\documentclass[12pt]{article}
\usepackage[margin=0.95 in]{geometry}
\usepackage{amsmath}
\usepackage{amssymb,amsfonts}
\usepackage[all]{xy}
\usepackage{graphicx}
\usepackage[utf8x]{inputenc}
\usepackage{amsmath}
\usepackage{amssymb}
\usepackage{float}
\usepackage{array}
\usepackage{tikz}
\usepackage{mathtools}
\usepackage{mathrsfs} 
\usepackage[hidelinks]{hyperref}
\usepackage{cite}
\usepackage{tikz-cd}
\numberwithin{equation}{section}
\numberwithin{equation}{section}
\numberwithin{table}{section}
\newcommand{\halg}{\mathfrak{h}}
\newcommand{\galg}{\mathfrak{g}}

\begin{document}

\thispagestyle{empty}

	\vspace*{3cm}
	{}
	
	\noindent
	{\LARGE \bf  Lie Algebra Fermions}
	\vskip .4cm
	\noindent
	\linethickness{.06cm}
	\line(10,0){467}
	\vskip 1.1cm
	\noindent
	\noindent
	{\large \bf  Jan Troost}
\vskip 0.25cm
{\em 
\noindent
 Laboratoire de Physique de l'\'Ecole Normale Sup\'erieure \\ 
 \hskip -.05cm
CNRS,  ENS, Universit\'e PSL, Sorbonne Universit\'e, Universit\'e de Paris,  Paris, France
}
	\vskip 1.2cm

	\vskip0cm
	
	\noindent {\sc Abstract: }    
	We define a supersymmetric quantum mechanics  of fermions that take values in a simple Lie algebra $\mathfrak{g}$. We summarize what is known about the spectrum and eigenspaces of the Laplacian which corresponds to the Koszul differential $d$.   Firstly, we  concentrate on the  zero eigenvalue eigenspace which coincides with the Lie algebra cohomology.  We provide physical insight into useful tools to compute the cohomology, namely Morse theory and the Hochschild-Serre spectral sequence. 
	We list explicit generators for the Lie algebra cohomology ring. Secondly, we concentrate on the eigenspaces of the supersymmetric quantum mechanics with maximal eigenvalue  at given fermion number. These eigenspaces have an explicit description  in terms of abelian ideals of a Borel subalgebra of the simple Lie algebra. We also introduce a model of Lie algebra valued fermions in two dimensions, where the spaces of maximal eigenvalue acquire a cohomological interpretation.	 Our work provides physical interpretations of results by mathematicians, and simplifies the proof of a few theorems. Moreover, we recall that these mathematical results  play a  role in pure	 supersymmetric gauge theory in four dimensions, and observe that they give rise to a canonical representation of the four-dimensional chiral ring.

	\vskip 1cm

	\pagebreak
	
	\newpage
	\setcounter{tocdepth}{2}
	\tableofcontents

	\section{Introduction}
	In this paper, we study  models of fermions that take values in a simple Lie algebra $\mathfrak{g}$. The quantization of the fermions gives rise to a Hilbert space equal to the Grassmann or exterior algebra $\Lambda \mathfrak{g}$  associated to the Lie algebra $\mathfrak{g}$.	
	There are a large number of beautiful results on the exterior algebra  of simple Lie algebras. In particular, the analysis of the spectrum of the Laplacian on the exterior algebra  \cite{KostantEigenvalues}, as well as the determination of the eigenspaces corresponding to the elements of maximal eigenvalue at given fermion degree has given rise to impressive theorems \cite{KostantLie,KostantEigenvalues,KostantOn,KostantPowers,KostantClifford}. 
	These go beyond the classic calculation of the Lie algebra cohomology \cite{Cartan,Koszul,CE,Reeder}.
	We translate these mathematical theorems into results on fermions in one and two dimensions that take values in the simple Lie algebra. We transcribe the mathematical operations and proofs into manipulations much more familiar to physicists. In a few instances, the physical translation   leads to  simpler  proofs.
	
	Besides the  motivation of linking up theoretical physics with mathematics and vice versa, there are  several intrinsically physical motivations to undertake this work. Firstly, our main model is a simple yet intriguing example of supersymmetric quantum mechanics. It is a system made of fermions only and with a simple Lie algebra symmetry. Yet, the spectrum and its group theoretic properties are highly non-trivial to compute and to a degree, unknown.
	Secondly, we make a link to the interesting subject of Lie algebra cohomology \cite{Cartan,Koszul,CE,Reeder}. An intricate version of Lie algebra cohomology plays a role in the analysis of the chiral ring of $N=2$ supersymmetric conformal field theories in two dimensions \cite{Lerche:1989uy,Hosono:1990yj}. We provide a  simpler application of Lie algebra 
	cohomology that makes  its wide applicability manifest.  Moreover, we describe an interesting relation of the supersymmetric quantum mechanics to  $N=1$ supersymmetric conformal field theories in two dimensions. Lastly, a subalgebra of the exterior algebra plays a crucial role in the conjectured description of the chiral ring of pure ${\cal N}=1$ supersymmetric Yang-Mills theories in four dimensions \cite{Cachazo:2002ry,Witten:2003ye,Etingof:2003dd,Kumar2,Kumar3,Etingof:2003fy}. More generally, the mathematics that we render more accessible applies to a degree to all quantum mechanical and field theoretical  models with Lie algebra valued fermions.
	
	The plan of the paper is as follows. In section \ref{sqm}, we present the supersymmetric quantum mechanics of simple Lie algebra valued fermions, including its Hilbert space, supercharges and Hamiltonian. Mathematical results on the spectrum of the Hamiltonian are  recalled in section \ref{eigenvalues}. 
	The eigenspace of zero eigenvalue is identified as the Lie algebra cohomology. We recall the calculation of the Lie algebra cohomology and explicit generators of the ring in section \ref{LieAlgebraCohomology}. We take the occasion to review a few essential ingredients in the proofs, like  Morse theory on the coset $G/H$ where $G$ is a compact connected group with simple
	Lie algebra $\mathfrak{g}$ and $H$ is a maximal torus of $G$, as well as the Hochschild-Serre spectral sequence. An attempt is made to make the mathematics transparent to physicists.
	In section \ref{maximal}, we  then concentrate on a subspace of the Hilbert space of the supersymmetric quantum mechanics  determined by the eigenspaces of maximal eigenvalue at given fermion degree. We review and simplify proofs of the results that describe the  subspace explicitly. In section \ref{explicit} we  provide a detailed 
	description of the subspace, summarizing a large body of mathematical literature on the subject.  A two-dimensional model of Lie algebra fermions is introduced in section \ref{twodimensions}.
	It provides  the physical backdrop for a cohomological interpretation of the  subspace, and allows to simplify the proof of a mathematical theorem.  In section \ref{SYM} we review the relevance of our model to the conjectured chiral ring of pure supersymmetric Yang-Mills theory in four dimensions, and observe that the uniform 	approach gives rise to canonical representatives of the chiral ring.
	Section \ref{conclusions} contains our conclusions. The appendices are devoted to technical details, longer calculations, or subjects that are slightly tangential to the main line of presentation.
	Throughout, we illustrate  general mathematical theorems with  concrete examples. 
	
	\section{The  Quantum Mechanics Of Lie Algebra Fermions}
	\label{sqm}

	\label{HilbertSpace}
	In this  section we first review the quantization of fermions that take values in vector spaces. We describe how quantization gives rise to the Grassmannian or exterior algebra of the vector space, and how standard operations on the exterior algebra translate into  physics language.
	We then apply the formalism to fermions taking values in a complexified simple Lie algebra $\mathfrak{g}$.
	The Lie algebra differential $d$ will provide a supercharge on the exterior algebra which defines a Hamiltonian equal to the Laplacian on the Lie algebra. The determination of the spectrum of the Hamiltonian and the irreducible representation content of the eigenspaces is a partially open problem. We  briefly review what is known about the spectrum. In the following sections, we will describe the eigenspaces of minimal and of maximal eigenvalue in detail.

	\subsection{The Exterior Algebra}
	Consider fermions $\psi$ taking values in a  vector space $V$, and dual fermions $\psi^\vee$ in the dual vector space	$V^\vee$.
	The fermions $\psi: \mathbb{R} \rightarrow V$ are maps from a time direction $\mathbb{R}$ into the vector space and similarly for the dual fermions.
	We introduce the free fermion action $S_{free}$:
	\begin{equation}
	S_{free} = \int_{\mathbb{R}} dt  ( \psi^\vee(t) , \partial_t \psi(t) )_{\text{eval}} \, . \label{firstaction}
	\end{equation}
	The bracket $(\cdot,\cdot)_{\text{eval}}$ equals the evaluation of the dual fermion $\psi^\vee$ on the derivative of the fermion $\psi$. The canonical quantization of the system gives
	rise to the anti-commutator relation
	\begin{equation}
	\{ \psi^\vee_a, \psi^b \} = \delta_a^b \label{canonicalanticommutator}
	\end{equation}
	where the fermions $\psi_a^\vee$ and $\psi^b$ are elements of dual bases of the vector spaces $V^\vee$ and $V$ respectively.	A representation space on which these fermion creation operators $\psi^b$ and
	annihilation operators $\psi^\vee_a$ act is the Grassmann or exterior algebra $\Lambda V$ of the vector space $V$. This is the space spanned by the anti-symmetrized tensor product of basis vectors and it has dimension $2^n$ where $n$ is the dimension of the vector space $V$. Formally, it is defined to be the tensor algebra divided by the two-sided ideal generated by vectors of the form $\psi \otimes \psi$ for $\psi$ a vector in the vector space $V$. We can view the vectors in the exterior algebra as linear combinations of wedge products of basis vectors. The exterior algebra $\Lambda V$ comes equipped with a few standard operations that include exterior and interior multiplication. In our physics language, and in the specific bases that we introduced, we have that
	the operators $\psi^b$ act as exterior multiplication by a basis element
	of $V$ while the operators $\psi^\vee_a$ act as interior multiplication by the basis element $\psi^a$. The canonical anti-commutation relation (\ref{canonicalanticommutator}) then follows from the relation between exterior and interior multiplication in the exterior algebra $\Lambda V$. 
	The anti-commutation relation  (\ref{canonicalanticommutator}) acquires the interpretation of a (Grassmannian) symplectic structure.
	The construction is very generic.
	
	\subsection{The Hilbert Space}	
	\label{GeneralQuantization}
	The quantization of the vector valued fermions  has given rise to a vector space $\Lambda V$, which doubles as an algebra.
	To obtain a Hilbert space for our quantum mechanics, we need a positive definite inner product in a complex vector space.  A simple starting point is to take the original vector space $V$ to be complex, namely, to consider complex fermions. We also consider the exterior algebra $\Lambda V$  to be defined over the complex numbers.  We define complex conjugation on $\Lambda V$ in terms of complex conjugation $\ast$ on the vector space $V$ by the formula $(\psi_1 \wedge \psi_2)^\ast=\psi_1^\ast \wedge \psi_2^\ast$. To define a positive definite inner product, we assume a symmetric bilinear product on the vector space $V$, denoted by $(\cdot,\cdot)_\kappa$.
	We define a corresponding inverse metric $\kappa^{ab}$
	\begin{equation}
	(\psi^a,\psi^b)_\kappa = \kappa^{ab}
	\end{equation}
	that we assume to be non-degenerate.
	The symmetric bilinear product allows us to identify the dual vector space $V^\vee$ with the vector space $V$, for instance through the bases identification:
	\begin{equation}
	(\kappa_{ab} \psi^b,\cdot)_\kappa = \psi^\vee_a \, ,
	\end{equation}
	where the metric matrix $\kappa_{ab}$ is the inverse of the matrix $\kappa^{ab}$.  Importantly, we assume that the basis $\psi^a$ is a basis of real vectors for which the metric $\kappa_{ab}$ is positive definite. We can then finally define  a positive definite inner product $( \cdot , \cdot )$ on the complex vector space $V$  by:
	\begin{equation}
	( c^1_a \psi^a, c^2_b \psi^b ) =  (c^1_a)^\ast c_b^2 (\psi^a, \psi^b)_\kappa=  (c^1_a)^\ast c^2_b  \kappa^{ab}  = (\psi^\ast_1,\psi_2)_\kappa \, .
	\end{equation}
	The positive definite inner product extends to $\Lambda V$ through linearity and the definition 
	\begin{equation}
	( \psi_1 \wedge \dots \wedge \psi_k , \chi_1 \wedge \dots \wedge \chi_k) = det ( \psi_i, \chi_j ) \, .
	\end{equation}
	Thus, the space $\Lambda V$ becomes a Hilbert space. We have quantized the action (\ref{firstaction}) and represented the canonical anti-commutator (\ref{canonicalanticommutator}) in a Hilbert space, under the assumption of the existence of a positive definite inner product on the underlying complex vector space $V$.
	\subsubsection*{Side Remark: The Clifford Algebra}
	There is an alternative model of fermions in a real vector space that is constructed as follows.
	We assume  that we have a  real vector space $V$ with a non-degenerate symmetric
	bilinear form $\kappa$. We define the action:
	\begin{equation}
	S_{Clifford} = \int dt (\psi,\partial_t \psi)_\kappa \, . \label{CliffordAction}
	\end{equation}
	Canonical quantization leads to the operator anti-commutator
	\begin{equation}
	\{ \psi^a , \psi^b \} = \kappa^{ab} \label{CliffordProduct}
	\, .
	\end{equation}
	The exterior algebra equipped with the Clifford product (\ref{CliffordProduct}) between the  fermions $\psi$ is a quantization of this model (that can be provided with a positive definite inner product under appropriate circumstances). We will mostly work with  the model of complex fermions described above.

	\subsection{Lie Algebra Valued Fermions}
	
	In this subsection, we apply the general formalism reviewed in subsection \ref{GeneralQuantization} to Lie algebra valued fermions. See also \cite{Koszul,KostantLie}.
	Consider a  complex simple Lie algebra $ {\mathfrak g}$ which is a complex vector space $V$.  The Lie algebra can be described as a direct sum of two real vector spaces  ${\mathfrak g}={\mathfrak q}+{\mathfrak t}$ 
	where ${\mathfrak t}$ is the Lie algebra of a compact Lie group. In other words, the Killing form $\kappa$ is positive definite on ${\mathfrak q}=i{\mathfrak t}$.\footnote{See Appendix \ref{Killing} for a review of the definition and a few properties of the symmetric bilinear Killing form.}  We choose
	a basis $\psi^a$ of the real vector space ${\mathfrak q}$ as well as the complex vector space ${\mathfrak g}$ and define complex conjugation $\ast$ in the algebra $\mathfrak g$ such that
	$(\psi^a)^\ast = \psi^a$. The structure constants ${f^{ab}}_{c}$ in the relation $[\psi^a,\psi^b] = {f^{ab}}_{c} \psi^c$
	are then imaginary since we have the commutator $[{\mathfrak q},{\mathfrak q}] =  i {\mathfrak q}$. 
	
As reviewed above, the Hilbert space of the supersymmetric quantum mechanics of fermions taking values in the complex simple Lie algebra $\mathfrak{g}$ is the exterior algebra $\Lambda \mathfrak{g}$. More explicitly, we denote the states in the Hilbert space by
	$|\Psi \rangle $ where
	\begin{equation}
	| \Psi \rangle =	\sum_{k=0}^{\text{dim} \, {\mathfrak g}} c_{a_1 \dots a_k} \psi^{a_1}  \wedge \dots \wedge \psi^{a_k} | 0 \rangle \, , \label{states}
	\end{equation}
	and the coefficients are anti-symmetric in their indices. The dimension of the Hilbert space is $2^{\text{dim} \, \mathfrak{g}}$.
	 We consider the variables $\psi^a$ to be anti-commuting (i.e. Grassmann) variables and will often omit the wedge symbols between the wedge product of fermions $\psi^a$. In the notation (\ref{states}), we have explicitly written a vacuum vector  $|0 \rangle $ on the far right in order to stress that these are vectors in a Hilbert space. This notation is  redundant and we  also omit it regularly in the following.
 We define ket states which represent the dual Hilbert space states:
	\begin{equation} 
	\langle 0 |	\sum_{k=0}^{\text{dim} \, {\mathfrak g}} c_{a_1 \dots a_k}^\ast \psi^{\dagger a_k} \wedge \dots \wedge \psi^{\dagger a_1} = (| \Psi\rangle)^\dagger \, . \label{dualstates}
	\end{equation}
	These dual states are defined such that the norm of the vacuum state is one, and the vacuum is annihilated by the operators
	$\psi^{\dagger a}$. We have the anti-commutation relation between annihilation and creation operators:
	\begin{equation}
	\{ \psi_a^\dagger, \psi^b \} = \delta_a^b \, , \label{anti-commutator} \, .
	\end{equation}
	We lower and raise indices on the fermions using the Killing metric $\kappa_{ab}$.\footnote{	
		Using the  Killing form, we  identify the dual Hilbert space with the Hilbert space:
		\begin{equation}
		(\psi^a,\psi^b) = \kappa^{ab} \, , \qquad (\psi^a, \cdot ) = \kappa^{ab} \langle 0 | \psi_b^\dagger = \langle 0 | 
		\psi^{\dagger a}\, .
		\end{equation}}
	\begin{equation}
	\langle 0 | \psi_a^\dagger \psi^b | 0 \rangle = \delta_a^b \, .
	\end{equation}
	As in the general case, we have the canonical exterior multiplication on the exterior algebra $\Lambda {\mathfrak g}$ 
	\begin{equation}
	\epsilon(\psi^a) = \psi^a \wedge = \psi^a \, .
	\end{equation}
	We also note that the interior multiplication $i(\psi_a)$ by a real basis element $\psi_a$ coincides with the hermitian conjugate  of the exterior multiplication $\epsilon(\psi^a)$.

	\begin{equation}
	i(\psi_a) = \psi^{\dagger}_a \, .
	\end{equation} 
	In summary, we raise and lower indices with the positive definite Killing metric (i.e. we identify the Lie algebra vector space with its dual using the inner product) and we have the operator translation between mathematics and physics:\footnote{The formulas are simpler and slightly less generic because we have chosen a real basis $\psi^a$.}
	\begin{eqnarray}
	\epsilon(\psi^a) & \leftrightarrow & \psi^a \, ,
	\nonumber \\
	i(\psi_a) & \leftrightarrow & \psi_a^\dagger \, .
	\end{eqnarray}

	\subsection{The Supersymmetric Quantum Mechanics}
	To identify an interesting supersymmetric quantum mechanics theory on the exterior algebra $\Lambda \mathfrak{g}$ of the simple Lie algebra $\mathfrak{g}$, we define a few operators \cite{Koszul}.
	Firstly, we define the operator $\theta(\psi^a)$ that implements the adjoint action of the basis element $\psi^a$ on the Lie algebra, extended to the exterior algebra:
	\begin{equation}
	\theta(\psi^a) =  {f^{ac}}_{b} \psi^b   \psi^\dagger_c \, .
	\end{equation}
	Indeed, the annihilation operator knocks out a fermion from a given state, to replace it with its commutator with $\psi^a$. The operator is quadratic in the fermions. Secondly, we introduce the derivative operator $d^\dagger$ in the fermion language and identify it with a supercharge $Q^\dagger$:
	\begin{equation}
	d^\dagger=\frac{1}{2} \epsilon( [\psi^a , \psi^b] ) i(\psi_b) i (\psi_a) 
	=\frac{1}{2} {f^{ab}}_c \psi^c \psi_b^\dagger \psi_a^\dagger = \frac{1}{2} \theta(\psi^a) i(\psi_a)= Q^\dagger
	\, . \label{definitionSupercharge}
	\end{equation}
	The operator $d^\dagger$ is cubic in the fermions and annihilates two fermions to replace them with their commutator. The invariant definition of the derivative $d^\dagger$  is \cite{Koszul}
	\begin{equation}
	d^\dagger(\psi_1 \wedge \dots \wedge \psi_p) = \sum_{i<j} (-1)^{i+j+1} [\psi_i,\psi_j] \wedge \psi_1 \dots \wedge \hat{\psi_i} \wedge \dots \wedge \hat{\psi_j} \wedge \dots \wedge \psi_p \, ,
	\end{equation} where all the factor elements $\psi_i$ are in the Lie algebra $ {\mathfrak g}$. It is straightforward to check that our  expression (\ref{definitionSupercharge}) does agree with the invariant definition.
	In equation (\ref{definitionSupercharge}), we introduced the  notation $d^\dagger=Q^\dagger$ that indicates that we consider the derivative operator  to be the supercharge $Q^\dagger$ of a supersymmetric quantum mechanics model.  The hermitian conjugate operator $Q$ equals 
	\begin{eqnarray}
	Q &=& - \frac{1}{2} {f_{ab}}^c \psi^a \psi^b \psi_c^\dagger = \frac{1}{2} \epsilon(\psi^a) \theta(\psi_a) =d \, ,
	\end{eqnarray}
	where we used the reality of the basis $\psi^a$ as well as the imaginary nature of the structure constants. Let us  introduce the operator
	$D$ that measures the fermionic degree of a state:
	\begin{eqnarray}
	D &=& \psi^a \psi_a^\dagger \, .
	\end{eqnarray}
	We	then  see that the operators  $Q$ and $Q^\dagger$ change the parity of the fermion degree $D$ and thus indeed can be interpreted as supercharges. In fact, the supercharge $Q^\dagger$ lowers the degree by one, and the supercharge $Q=d$ raises it by a unit.
	In standard fashion, we  define the supersymmetric quantum mechanics  Hamiltonian $H$:
	\begin{eqnarray}
	\frac{H}{2} &=& Q Q^\dagger + Q^\dagger Q \, . \label{definitionHamiltonian}
	\end{eqnarray}
	Thus, we are ready to study the supersymmetric quantum mechanics model living in the Hilbert space $\Lambda \mathfrak{g}$. 
	
	We first establish a number of elementary facts.
	We note that the supercharge commutes with the adjoint action
	\begin{equation}
	Q^\dagger \theta(\psi^a)=\theta(\psi^a) Q^\dagger \, ,
	\end{equation}
	and therefore so does the Hamiltonian $H$. Consequently, the eigenspaces of the Hamiltonian represent the adjoint
	action of the simple Lie algebra $\mathfrak{g}$.
	We moreover have handy formulas of the type \cite{Koszul}:
	\begin{equation}
	\epsilon(\psi^a) d^\dagger+ d^\dagger \epsilon(\psi^a) = \theta(\psi^a)
	\end{equation}
	which translates to the easily verified fermion operator identity
	\begin{equation}
	\psi^a   \frac{1}{2} {f^{db}}_c \psi^c \psi_b^\dagger \psi_d^\dagger
	+ \frac{1}{2} {f^{db}}_c \psi^c \psi_b^\dagger \psi_d^\dagger \psi^a =  {f^{ab}}_c \psi^c \psi_b^\dagger 
	= \theta(\psi^a) \, . \label{exteriorderivativecommutator}
	\end{equation}
	The anti-commutator (\ref{exteriorderivativecommutator}) is useful to simplify the Hamiltonian:	
	\begin{eqnarray}
	H &=& -\frac{1}{2} {f^a}_{bc} \psi_a \psi^b \psi^{\dagger,c}  {f^{de}}_g \psi^g \psi_e^\dagger \psi_d^\dagger 
	- \frac{1}{2} {f^{de}}_g \psi^g \psi_e^\dagger \psi_d^\dagger   {f^a}_{bc} \psi_a \psi^b \psi^{\dagger,c}
	\nonumber \\
	&=& -\frac{1}{2}  \psi_a    {f^{de}}_g \psi^g \psi_e^\dagger \psi_d^\dagger  {f^a}_{bc}  \psi^b \psi^{\dagger,c}
	- \frac{1}{2} {f^{de}}_g \psi^g \psi_e^\dagger \psi_d^\dagger  \psi_a  {f^a}_{bc}  \psi^b \psi^{\dagger,c}
	\nonumber \\
	&=& - {f^{}}_{aeg} \psi^g \psi^{\dagger,e}  {f^a}_{bc}  \psi^b \psi^{\dagger,c} =  \theta(\psi_a) \theta(\psi^a) 
	=  C_2(\theta) \label{HamiltonianCasimir}
	\, .
	\end{eqnarray}
	In summary, the Hamiltonian of the supersymmetric quantum mechanics equals the quadratic Casimir $C_2$ of the adjoint action $\theta$ on the Hilbert space $\Lambda \mathfrak{g}$, and the 
	eigenspaces represent the adjoint action of the Lie algebra.

	\section{Eigenvalues and Irreducible Eigenspaces}
	\label{eigenvalues}
	In the previous section, we defined a supersymmetric quantum mechanics on the Hilbert space $\Lambda {\mathfrak g}$ of a simple Lie algebra $\mathfrak g$.
	In this section, we compile information on the eigenspaces of the Hamiltonian $H$. We are also interested in the decomposition of the eigenspaces into irreducible representations of the adjoint  action of the Lie algebra on the exterior algebra $\Lambda \mathfrak{g}$. 
	Firstly, we provide example calculations at low rank. We then turn to a discussion of known generic characteristics of the solution. We  recall that the problem of determining the zero modes is solved by the Lie algebra cohomology. A review of the calculation of the Lie algebra cohomology is the subject of section \ref{LieAlgebraCohomology}. The maximal eigenvalue problem at given fermion number  will occupy us in the subsequent sections.

	\subsection{Low Rank Examples}
	To understand the problem of determining the eigenvalues of the Hamiltonian $H$, as well as the irreducible representations spanned by the eigenspaces, we find it instructive to provide a few low rank examples. These  also serve to illustrate the general theorems that we will enumerate shortly.
	
	\subsubsection{The Lie Algebra $\mathfrak{su}(2)$}	
	The simplest simple Lie algebra $\mathfrak{g}$ is the complexified $\mathfrak{su}(2)$ algebra. 
	The spectrum of the Hamiltonian on the Hilbert space is easily checked to be the following. We have a scalar $ | 0 \rangle$ at fermion degree zero,  an adjoint $\psi^a | 0 \rangle$ at fermion degree one, an adjoint at degree two (arising from the anti-symmetric product of two adjoints of $\mathfrak{su}(2)$) and a scalar at degree three. The dimension of the Hilbert space is eight (namely two to the power of the dimension of the Lie algebra). The scalars have zero Hamiltonian. A calculation, or standard formulas for the quadratic Casimir show that the Hamiltonian equals one at degree one and at degree two.  We can characterize the representation theoretic content at each degree through the polynomial $P(y) = 1 + 3 y + 3 y^2 + y^3$, where the coefficients provide the dimensions of the irreducible representations at the fermion degree given by the power of the variable $y$. The Hamiltonian is the quadratic Casimir evaluated on the irreducible representations. If we denote representations by their Dynkin label, then the characteristic polynomial $P(y)$ reads:
	\begin{equation}
	P(y) = (0) + (2) y + (2) y^2 + (0) y^3 \, . \label{characteristicpolynomialsu2}
	\end{equation}

	\subsubsection{The Lie Algebra $\mathfrak{su}(3)$}	
	More generally, the graded character $\chi(y)$ of the representation $\Lambda  {\mathfrak g}$ is given by the formula:
	\begin{equation}
	\chi(y) =  \prod_{\mu  \, \in \, \text{adjoint}} (1+y e^\mu) \, , \label{characteristicpolynomialg}
	\end{equation}
	where the product is over weights $\mu$ in the adjoint representation. Indeed, each fermion $\psi^a$ of weight $\mu$ is either present or not in a given state in $\Lambda \mathfrak{g}$.
	The  representation content and spectrum is obtained by a weight space decomposition, performed degree by degree.\footnote{This is elementary group theory. The algorithm is to identify the highest weight, subtract the corresponding irreducible character, and continue until done. One uses the decomposition theorem for finite dimensional representations of simple Lie algebras.}  We find  the resulting $\mathfrak{su}(3)$ characteristic polynomial $P(y)$
	\begin{eqnarray}
	P(y) &= & (0,0) + (1,1) y + ((3,0)+(0,3)+(1,1) ) y^2 
	\nonumber \\
	& & + ( (0,0)+(1,1)+(3,0)+(0,3)+(2,2))y^3 + (2 (1,1) + 2 (2,2) ) y^4 
	\nonumber \\
	& &+ ( (0,0)+(1,1)+(3,0)+(0,3)+(2,2))y^5 + ((3,0)+(0,3)+(1,1) ) y^6
	\nonumber \\
	& & 
	+ (1,1) y^7 
	+ (0,0) y^8  \, .  \label{characteristicpolynomialsu3}
	\end{eqnarray}
	We again labeled representations by their Dynkin labels $(d_1,d_2)$ which correspond to irreducible representations with highest weight  $\lambda = d_1 \lambda_1 + d_2 \lambda_2$ where $\lambda_{1,2}$ are the fundamental weights of $\mathfrak{su}(3)$.
	The dimension and quadratic Casimir  of these representations are:
	\begin{equation}
	\begin{array}{cccccc}
	(d_1,d_2) & (0,0) & (1,1) & (3,0) & (0,3) & (2,2) \\
	\text{dim} & 1 & 8 & 10 & 10' & 27 \\
	C_2 &	0  & 1 & 2 & 2 & 8/3 
	\end{array}
	\end{equation}
	For future reference, note that the Poincar\'e palindromic polynomial (\ref{characteristicpolynomialsu3})  implies that we have invariants at degrees zero, three, five and eight for a total of four invariants.	
	These elementary calculations can also be found in \cite{Reeder} along with a longer list of interesting examples.
	At this stage it should be clear that there is an algorithm for determining the spectrum of the Hamiltonian at each degree for each given example. Mathematicians have tried to describe the generic properties of the final result of the algorithm (to decompose  into irreducible representations and evaluate the quadratic Casimir) more succinctly.

	\subsection{The Zero Modes and The Cohomology of Lie Algebras}
	The eigenspace of the positive definite Hamiltonian $H$ corresponding to zero eigenvalue consists of states that must be annihilated by the supercharges $d=Q$ and $d^\dagger=Q^\dagger$ as well as the adjoint action $\theta$ (in view of equations (\ref{definitionHamiltonian}) and (\ref{HamiltonianCasimir})). Therefore, this is the space of harmonic forms, or invariants in the exterior algebra. The calculation of this eigenspace is known to be equivalent to the calculation of the de Rham cohomology with real coefficients on compact Lie groups,  which in turn is equivalent to the calculation of the Lie algebra cohomology \cite{CE}.
	The de Rham cohomology ring $H(G)$ of a simple, compact and connected Lie group $G$ was calculated in the second quarter of the twentieth century. See \cite{Samelson,BorelSurvey} for historical surveys with references. It  is the cohomology ring of a product of odd dimensional spheres. The final result is easy to comprehend and find, but the proof is harder and more difficult to reconstruct.
	The space of invariants in the exterior algebra is freely generated by  odd elements of order $2m_i+1$ where $m_i$ are the exponents of the Lie algebra (i.e. the order of the Casimir invariants of the Lie algebra, minus one).
	\begin{table}[h!]
		\centering
		\begin{tabular}{|c|c|}
			\hline 
			$ {\mathfrak g}$ & 
			 {Exponents} \\
			\hline
			$\mathfrak{a}_{r} $& 
			$1,2,\dots,r  $\\
			\hline 
			$\mathfrak{b}_{r} $& 
			 $1,3,5,\dots,2r-1 $\\
			\hline 
			$	\mathfrak{c}_{r} $& 
			$1,3,5,\dots, 2r-1$ \\
			\hline 
			$	\mathfrak{d}_{r} $& 
			$1,3,5, \dots, 2r-3,r-1$ \\
			\hline 
			$	\mathfrak{e}_6$ & 
			 $1,4,5,7,8,11$ \\
			\hline 
			$\mathfrak{e}_7 $& 
			$ 1,5,7,9,11,13,17$\\
			\hline 
			$	\mathfrak{e}_8 $& 
			 $1,7,11,13,17,19,23,29$\\
			\hline 
			$\mathfrak{f}_4$ & 
			$ 1,5,7,11 $\\
			\hline 
			$\mathfrak{g}_2$ &
			$ 1,5$\\
			\hline 
		\end{tabular}
		\caption{ Lie Algebra exponents and dual Coxeter numbers. 
		}
		\label{exponents}
	\end{table}
	Table \ref{exponents} provides the exponents  of all simple Lie algebras $\mathfrak{g}$.  There are $\mbox{rank} \, \mathfrak{g}=r$ such odd elements and therefore the cohomology ring has dimension $2^{\text{rank} \, \mathfrak{g}}$. The characteristic polynomial $P^{\mathfrak{g}}$ for the scalars in the exterior algebra is therefore:
	\begin{equation}
	P^{\mathfrak{g}}(y) = 	\sum_n dim (\Lambda^n \mathfrak{g})^{\mathfrak{g}} y^n =
	\prod_{i=1}^{\text{rank} \, \mathfrak{g}} (1+y^{2m_i+1}) \, . \label{invariants}
	\end{equation}
	For instance, for $\mathfrak{su}(3)$ we have that $m_1=1$ and $m_2=2$ which leads to the characteristic polynomial of invariants $P^{\mathfrak{g}}(y) = (1+y^3)(1+y^5)$ which agrees with the four invariants of degrees $0,3,5$ and $8$ in equation (\ref{characteristicpolynomialsu3}). Since sifting invariants out of a high dimensional representation through the generic reduction algorithm is challenging, the general Lie algebra cohomology result (\ref{invariants}) makes a  powerful point. Section \ref{LieAlgebraCohomology} is dedicated to clarifying this result.

	\subsection{The Adjoint}
	A second generic result is that the number of adjoint representations at each degree is  captured by the
	characteristic polynomial $P^{\text{adj}}(y)$ \cite{CPP}:	
	\begin{equation}
	P^{\text{adj}}(y) = \sum_n m^{\text{adj}} (\Lambda^n \mathfrak{g}) y^n = \prod_{i=1}^{\text{rank} \, \mathfrak{g}-1}(1+y^{2m_i+1}) \sum_{i=1}^{\text{rank} \, \mathfrak{g}}(y^{2m_i-1}+y^{2m_i}) \, ,
	\end{equation}
	where $m^{\text{adj}}(\Lambda^n \mathfrak{g})$ is the multiplicity of the adjoint representation at fermion degree $n$.
	For example, for the algebra $\mathfrak{su}(3)$ the adjoint polynomial equals $P^{\text{adj}} = (1+y^3) (y+y^2+y^3+y^4)$, again in agreement with the characteristic $\mathfrak{su}(3)$ polynomial (\ref{characteristicpolynomialsu3}).

	\subsection{Low Degree Examples} 
	Let us also  provide general results of a different nature. Firstly, note that at degree zero, there is a single scalar, and the Hamiltonian equals zero. We can also calculate the final answer for the subspace of fermionic degree equal to one.   At degree one, we have an adjoint representation, and our quadratic Casimir is normalized such that it equals one.  Thus the Hamiltonian equals one, which again equals the degree of the space. For illustrative purposes, let us also calculate this result explicitly:
	\begin{eqnarray}
	H \psi^a | 0 \rangle 
	&=& \theta(\psi_b) \theta(\psi^b) \psi^a | 0 \rangle
	\nonumber \\
	&=& \theta(\psi_b) {f^{ba}}_c \psi^c  | 0 \rangle
	\nonumber \\
	&=&  -{f^{ba}}_c {f^c}_{bd} \psi^d | 0 \rangle
	\nonumber \\
	&=& \psi^a | 0 \rangle \, , \label{degreeone}
	\end{eqnarray}
	where we used the Killing metric property ${f^{ba}}_c {f^c}_{bd}=-\delta^a_d$ reviewed in appendix \ref{Killing}.
	
	At degree two, we distinguish the subspace $Q \mathfrak{g} \subset \Lambda^2 \mathfrak{g}$ generated by the action of the supercharge $Q$ on the Lie algebra $\Lambda^1 \mathfrak{g}=\mathfrak{g}$.     
	It contains linear combinations of elements $ d\psi^a =-1/2 {f^a}_{bc} \psi^b \psi^c$. Consider a generic element $c_{1b} c_{2d} \psi^b  \psi^d$ of degree two, and note that its inner product with $d \psi^a $ is proportional to:
	\begin{equation}
	\langle 0 | c_{1}^{b \ast} c_{2}^{d\ast} \psi^{\dagger }_d \psi^{\dagger }_b  {f^a}_{gh} \psi^g \psi^h | 0 \rangle
	=  2 c_{1}^{b \ast} c_{2}^{d \ast} {f^a}_{bd} \, .
	\end{equation}	
	The generic state  is orthogonal to the full space $Q \mathfrak{g}$  if it consists of two commuting Lie algebra elements $\psi_{i} = c_{ib} \psi^b$. We can therefore orthogonally split $\Lambda^2 g = A_2 + Qg$ where $Qg$ is the space of $Q$ exact states and the space  $A_2$ is the space of commuting bi-vectors $\psi_1 \wedge \psi_2$. On the first space, the eigenvalue of the Hamiltonian is equal to one, while on the second it  equals  two. For illustration, let us show these properties explicitly. We compute:
	\begin{eqnarray}
	H  \psi^n \psi^m | 0 \rangle 
	&=& 	{f^a}_{bc} \psi^b \psi^{\dagger c} f_{ade} \psi^d \psi^{\dagger e} \psi^n \psi^m  | 0 \rangle
	\nonumber \\
	&=& 2 \psi^n \psi^m - {{f^a}_b}^d f_{amn} \psi^b \psi^d | 0 \rangle \, . \label{degreetwo}
	\end{eqnarray}
	When we multiply by $c_{1n} c_{2m}$ and suppose that $\psi_1$ and $\psi_2$ commute, the last term vanishes, and we obtain an eigenvalue equal to two. When we have an elements of $Q \mathfrak{g}$ at hand, we multiply by ${f^{c}}_{nm}$ and use the Killing metric property to simplify the last term, which subtracts one from the eigenvalue, to give a total of one. Thus, the degree two problem is solved by explicit calculation. It provides a glimpse of a more general structure that we shall explore from section \ref{maximal} onwards.

	\subsection{Further Remarks}
	There are other results, e.g. for multiplicities of representations of small highest weight inside the exterior algebra $\Lambda \mathfrak{g}$, as well as for representations with highest weight close to the maximal highest weight $2 \rho$ where $\rho$ is half the sum of all positive roots \cite{Reeder}.\footnote{The maximal highest weight follows directly from equation (\ref{characteristicpolynomialg}).} We also know that the total representation equals $2^{\text{rank} \, \mathfrak{g}}$ times the tensor product of two irreducible representations with highest weight $\rho$ \cite{KostantClifford,KostantEigenvalues}.
	However, the question of the determination of the  multiplicities of all irreducible representations  awaits a  closed formula answer.
	
	In the rest of this paper, we will be concerned with studying a particular subset of eigenspaces, namely those that have zero eigenvalue and those that have maximal eigenvalue at a given fermion degree. In these cases, a more explicit description of the eigenspaces can be obtained. The answers are  relevant to physical theories  beyond the supersymmetric quantum mechanics model at hand.

	\section{Gauged Lie Algebra Fermions}
	\label{GaugedLieAlgebraFermions}
	\label{GroupCohomology}
	\label{LieAlgebraCohomology}
	
	In section \ref{sqm} we defined a supersymmetric quantum mechanics whose zero eigenvalue eigenvectors are the invariants in the exterior algebra of the Lie algebra.
	There is a second model that gives rise to the same space of states. Before we study the space, we introduce this second model.
	Firstly, we define a  group $G$ which is a connected and compact exponentiation of the real Lie simple algebra $\mathfrak{t}$. The group $G$ acts by the adjoint action on the Lie algebra (as well as its complexification $\mathfrak{g}$), and by extension it acts on the exterior algebra. We can gauge the adjoint action. The space of gauge invariants then coincides with the zero eigenvalue eigenspace of the supersymmetric quantum mechanics introduced in section \ref{sqm}. Thus, we have a second model for the eigenspace.
	
	The space of invariants is denoted $(\Lambda \mathfrak{g})^G$ or $(\Lambda \mathfrak{g})^{\mathfrak{g}}$. It is a classic result that the de Rham cohomology ring $H(G)$ with real coefficients equals the ring of gauge invariants $(\Lambda \mathfrak{g})^G$ \cite{CE}. Thus, the question of providing a list of gauge invariants in this  simple model of gauged quantum mechanics is equivalent to computing the cohomology ring $H(G)$ of the simple Lie group $G$. Since the latter is not straightforward to calculate, we succinctly review the literature. The reader only interested in the maximal eigenvalue subspaces of the supersymmetric quantum mechanics model of section \ref{sqm} can safely skip to section \ref{maximal}. 
	
	This long section is subdivided as follows. In subsection \ref{GaugedModel} we describe the gauged fermion model in detail. The number of gauge invariants states is computed in subsection \ref{NumberOfStates}.  How to calculate explicit representatives is summarized in subsection \ref{ExplicitInvariants}. In subsection \ref{Solution} we review one manner of computing the group cohomology. The  cohomology of
	the  geometric coset space $G/H$ where $H$ is a maximal torus of the group $G$ enters the calculation. 
	This  leads to the introduction of a second model, which is the quantum mechanics of a superparticle on the coset space $G/H$, which we discuss in more detail in appendix \ref{Coset}. Our main references for this section are \cite{Reeder,Solleveld}.

	\subsection{The Gauged Lie Algebra Fermions}
	\label{GaugedModel}
	\label{GaugedFermions}
	
	We consider again the Lie algebra valued fermions $\psi^a$ of section \ref{sqm}. We gauge the
	adjoint action  $G \times \mathfrak{g} \rightarrow \mathfrak{g} : (g,\psi) \mapsto g \psi g^{-1}$ of the group $G$ on the Lie algebra $\mathfrak{g}$. 
	The invariance of the Killing metric $\kappa$ under the adjoint action ensures that we can gauge the symmetry group and formulate the simple interacting model of gauged Lie algebra fermions:
	\begin{eqnarray}
	S_{gauged} &=& \int_{L} d t ( \psi^\ast,  \partial_t \psi - [A_t,\psi])_\kappa  \,  , \label{GaugedAction}
	\end{eqnarray}
	where we introduced the Lie algebra valued gauge field one form $A$ which we pulled back to the world line $L$ of our fermions. After quantization, the full Hilbert space is still the exterior algebra $\Lambda \mathfrak{g}$. The gauge invariant states are all the states that are invariant under the adjoint action of the gauge group $G$, and they span the space of physical states $(\Lambda \mathfrak{g})^G$ in this model. It is the latter space that we wish to discuss in detail in this section. Firstly, we  compute its dimension.

	\subsection{The Number of States}
	\label{NumberOfStates}

	After quantization of the action (\ref{GaugedAction}), the Gauss constraint implies that the operators
	\begin{equation}
	\theta (\psi^a) = {f^{ac}}_{b}  \psi^b \psi^{\dagger}_c
	\end{equation}
	that implement the adjoint action of the gauge group
	must annihilate physical states for all values of the adjoint index $a$. This is the Hamiltonian implementation of gauge invariance. 
	The Lagrangian path integral realization is also interesting and leads more directly to a counting of the number of gauge invariant states. Indeed, let us consider the world line $L$ of the fermions to be a circle. The path integral 
	\begin{equation}
	Z = \int dA \int d\psi d \psi^\ast \exp (S_{gauged}[A,\psi,\psi^\ast])
	\end{equation}
	over circle maps then computes the trace over the Hilbert space. The integral over the time component of the gauge field implements gauge invariance. Let us study the path integral in more detail. 
	
	 The path integral over gauge connections is an integral over principal $G$-bundles over the circle. Since these are bundles over a one-dimensional manifold, they are  flat. Principal bundles with flat connection over the circle are classified by their monodromy, up to conjugation (see e.g. \cite{Taubes}). Thus, we can replace the path integral over the gauge connection by an integral over the compact group manifold with conjugation invariant measure. In much more detail, the same result can be obtained by first choosing Lorenz gauge $\partial_t A_t=0$, 
	then implementing the Faddeev-Popov procedure, thirdly, integrating over constant gauge connections $A_t=a$, and finally proving that the combination of the Faddeev-Popov determinant and the measure over the Lie algebra leads to the Haar measure on the group. For instances of detailed derivations in the physics literature see. e.g. \cite{Reinhardt:1996fs,Aharony:2003sx}.
	The upshot is that one finds the path integral simplification:
	\begin{eqnarray}
	Z &=& \int d_{H} a  \int d \psi d \psi^\ast  e^{S_{free}}  e^{ [a,\psi] \psi^\ast} 
	\end{eqnarray}
	We have denoted the Haar measure by $d_H a$. The quantization of the fermions gives rise to the Hilbert space $\Lambda \mathfrak{g}$ over which we trace and the operator we trace is the adjoint action by the constant gauge field $a$:
	\begin{eqnarray}
	Z
	&=&
	\int d_{H} a  Tr  e^{\text{ad} (a)} =\int d_{H} a  \,  Tr \text{Ad} (e^{a}) 
	\end{eqnarray}
	We used that the exponential of the adjoint action on the Lie algebra is the adjoint action of the exponential. Finally,  we can rewrite  the measure in terms of the (invariant Haar) measure $dg$ on the group manifold. We moreover use that the operator $\text{Ad} g$ on the Lie algebra $\mathfrak{g}$  with eigenvalues $\lambda_a$
	has eigenvalues $\prod_a (1+\lambda_a)=\det(1+\text{Ad} g)$ on $\Lambda \mathfrak{g}$ when extended to the exterior algebra, and find:
	\begin{eqnarray}
	Z
	&=&\int_G d g  \det (1+\text{Ad} g) \, . \label{integraldeterminant}
	\end{eqnarray}
	We can further compute the integral (\ref{integraldeterminant}) by the Weyl integration formula on the group that reduces the integral of the class function to an integral over the Cartan torus $H$. We evaluate
	the function on the Cartan torus and add the appropriate measure factor $\det(1-\text{Ad} \, h)|_{\mathfrak{m} } $ where $\mathfrak{m}$ is the orthogonal complement of the Cartan subalgebra $\mathfrak{h}$ inside the (real compact) Lie algebra $\mathfrak{t} = \mathfrak{h} \oplus \mathfrak{m} $. We also divide by the cardinal number $|W|$ of the Weyl group, as dictated by the integration formula:
	\begin{eqnarray}
	Z &=& \frac{1}{|W|} \int_H dh \det (1+\text{Ad} \, h) \det(1-\text{Ad} \, h)|_{\mathfrak{m} } = \frac{2^{{\text{rank} \, \mathfrak{g}}}}{|W|} \int_H  dh \det (1-\text{Ad} \, h^2)|_{\mathfrak{m}} 
	\nonumber \\
	&=& \frac{2^{{\text{rank} \, \mathfrak{g}}}}{|W|} \int_H dh \det (1-\text{Ad}(h))|_{\mathfrak{m}} 
	= 2^{\text{rank} \, \mathfrak{g}} \int_G dg = 2^{\text{rank} \, \mathfrak{g}}  \, ,
	\end{eqnarray}
	where ${\text{rank} \, \mathfrak{g}} $ is the rank of the group $G$ and the Haar measure is properly normalized. Thus, the dimension of the space of gauge invariants is equal to $2^{\text{rank} \, \mathfrak{g}} $.

	\subsection{Explicit Invariants}
	\label{ExplicitInvariants}
	Generic explicit representatives can be constructed on the basis of the invariant polynomials taking values in the Lie algebra and Cartan's theory of transgression \cite{CartanNotions}. This fact is very briefly summarized in \cite{PS}.
	For the classic infinite series of Lie algebras explicit representatives of the group invariants can be formulated in terms of traces and determinants	\cite{GHV}.
	The generic prescription to construct invariants from polynomials is to take a group invariant polynomial of degree $d_i=m_i+1$, and to replace $d_i-1$ elements by commutators, and one by an ordinary Lie algebra element, and anti-symmetrize over all $2 d_i-1$ entries. These are then group invariant representatives of the Lie algebra cohomology. For example, the quadratic invariant polynomial gives rise to the generic cubic Lie algebra cohomology element 
	\begin{equation}
	I_3^{\text{generic}} = \kappa^{ad} {f_d}^{bc} \psi_a \wedge \psi_b  \wedge \psi_c \, . \label{cubicinvariant}
	\end{equation}
	Below we illustrate  the generic prescription in a few low rank examples.
	\subsubsection{$\mathfrak{su}(2)$}
We saw that in the exterior algebra $\Lambda \mathfrak{su}(2)$ the invariants live at degree zero and degree three. Thus, they are easily written down with the knowledge at hand:
	\begin{equation}
	I_0 = 1 \, , \qquad  \qquad  I_3 = \psi^\alpha \wedge \psi^{-\alpha} \wedge \psi^{H} \, ,
	\end{equation}
	where $\psi^{\pm \alpha}$ correspond to the generators $E^{\pm \alpha}$ and $\psi^{H}$ lies in the Cartan direction. See appendix \ref{ExplicitAlgebras} for our  conventions. The invariant at degree zero is trivial, and the invariant
	at degree three is proportional to the invariant (\ref{cubicinvariant}).
	
	\subsubsection{$\mathfrak{su}(3)$}

A basis for the Lie algebra $\mathfrak{su}(3)$ is made up of two Cartan generators $H_1,H_2$ as well as the root vectors $E_{\pm \alpha_{1,2}}, E_{\pm \theta}$
where $\theta=\alpha_1+\alpha_2$ is the highest root. Their non-trivial commutation relations, inner product with respect to the Killing form and other properties are recorded
in appendix \ref{ExplicitAlgebras}. 
	We denote the corresponding elements of the algebra $\Lambda^1 \mathfrak{g}$ by $\psi_{1,2}$ and $\psi_{\pm \alpha_{1,2},\pm \theta}$.
	There are four invariants, at degree zero, three, five and eight. 
	The invariants at degree zero and eight are  obvious:
	\begin{equation}
	I_0= 1 \, , \qquad \qquad  I_8 = \prod_{ \mu  \in \text{adj}}  \psi^{\mu} \, .
	\end{equation}
	The invariants at degree three and five are:
	\begin{eqnarray}
	I_3 &=&   \psi_{\alpha_1} \psi_{\alpha_2} \psi_{-\theta} 
	- \psi_{-\alpha_1} \psi_{-\alpha_2} \psi_{\theta}
	+  \psi_{\alpha_1} \psi_{-\alpha_1} \psi_1
	+ \psi_1  \psi_{\theta} \psi_{-\theta}
	+ \psi_{\alpha_2} \psi_{-\alpha_2} \psi_{2}
	+ \psi_2 \psi_{\theta} \psi_{-\theta}  \, ,  \label{su3invariants}
	 \\
	I_5 &=&    -2 \psi_{2} \psi_{\theta} \psi_{\alpha_2} \psi_{-\theta} \psi_{-\alpha_2}
	- \psi_{1} \psi_{\theta} \psi_{\alpha_2} \psi_{-\theta} \psi_{-\alpha_2}
	+  \psi_{2} \psi_{\alpha_1} \psi_{-\alpha_1}\psi_{\alpha_2}  \psi_{-\alpha_2}
	-  \psi_{1} \psi_{\alpha_1} \psi_{-\alpha_1}\psi_{\alpha_2}  \psi_{-\alpha_2} \nonumber  \\
	& &
	+ \psi_{1} \psi_{2} \psi_{\theta} \psi_{-\alpha_1} \psi_{-\alpha_2}
	+ \psi_{1} \psi_{2} \psi_{\alpha_1} \psi_{\alpha_2} \psi_{-\theta} 
	+2 \psi_{1} \psi_{\alpha_1} \psi_{\theta} \psi_{-\alpha_1} \psi_{-\theta}
	+ \psi_2 \psi_{\alpha_1} \psi_{\theta} \psi_{-\alpha_1} \psi_{-\theta} \, . \nonumber 
	\end{eqnarray}
	One can obtain them through a brute force calculation. Alternatively, the invariant at degree three  is proportional to the expression (\ref{cubicinvariant}). The quintic invariant agrees with contracting the cubic Casimir $d^{abc}$ with two commutators, $I_5 =
	d^{abc} {f^{de}}_a {f^{gh}}_b \psi_d \psi_e \psi_g \psi_h \psi_c$, again following the generic prescription.
Moreover, the quintic invariant is consistent with Poincar\'e duality (i.e. it can be related to the cubic invariant by contracting with the fully anti-symmetric tensor with appropriately raised and lowered indices).

Of course, the examples agree with the generic statements made earlier about their total number as well as their degree. One also notes that their explicit form rapidly becomes complicated to compute by brute force. Therefore the  construction for generic representatives is useful. The proofs of the completeness of the basis provided by the  prescription above are either through Cartan's transgression \cite{CartanNotions}, or they proceed through the calculation of a relative Lie algebra cohomology. We sketch the latter path in the next subsection, following \cite{Reeder,Solleveld}.

	\subsection{The Cohomology}
	\label{Solution}
	\label{TheCohomology}
	\label{Cohomology}  
	In subsection \ref{NumberOfStates} we determined that the dimension of the space of gauge invariants in the Hilbert space of Lie algebra valued fermions equals $2^{\text{rank} \, \mathfrak{g}}$. In this subsection, we calculate the cohomology more explicitly  by first identifying a Cartan subalgebra $\halg$ of the simple Lie algebra $\galg$ and then splitting the calculation of the cohomology into three steps. The three steps are the first and most crucial stages in what is known as the Hochschild-Serre spectral sequence \cite{HochschildSerre}. We review these steps in slightly more elementary terms.

	\subsubsection{The Spectral Sequence}
	
	Consider a Cartan subalgebra $\halg$ of the complex simple Lie algebra $\galg$. We analyse  how the differential $d$ on $\Lambda \galg^\ast$ acts with respect to the decomposition 
	$\Lambda \galg^\ast = \Lambda \halg^\ast \otimes \Lambda (\galg/\halg)^\ast$ where we think of the quotient space $\mathfrak{m}=\galg/\halg$ as the orthogonal complement of $\halg$ with respect to the Killing metric.\footnote{The exterior algebra $\Lambda \mathfrak{g}^\ast$ is canonically isomorphic to $\Lambda \mathfrak{g}$. The use of the dual Lie algebra in the present paragraph more easily allows for generalizations beyond simple Lie algebras.} The decomposition in terms of the algebra $\Lambda \halg^\ast$ and its complement provides a second grading to the space $\Lambda \galg^\ast$. On the one hand, we have the total degree of a (homogeneous) form and on the other hand we have  the degree that counts the number of factors that take values in the  subalgebra $\halg^\ast$. The main observation is that we can decompose the differential $d$ according to these two degrees. Firstly, all terms in the differential augment the total degree by one. Secondly, there are terms in $d$ that change the $\halg^\ast$ degree  by one, by zero or by minus one.   This is a consequence of the fact that $[ \mathfrak{m} , \mathfrak{m} ] \subset \mathfrak{m} \oplus \mathfrak{h}$ and $[ \halg , \mathfrak{m} ] \subset \mathfrak{m}$.
	
	There is a mathematical construction that exploits these different behaviors of the differential in order to approximate the cohomology. This is the Hochschild-Serre spectral sequence \cite{HochschildSerre}. The spectral sequence approximates the cohomology in multiple steps. We study the first three steps that correspond to the triple split of the differential in terms of $\halg^\ast$ degree.
	
	The construction of the spectral sequence proceeds as follows. Consider the spaces $F^p \Lambda^k (\mathfrak{g}^\ast)$ which are defined as follows
	\begin{equation}
	F^p \Lambda^k (\mathfrak{g}^\ast)=\{ \Psi \in \Lambda^k (\mathfrak{g}^\ast): \forall \psi_i \in \halg, i(\psi_1) \dots i(\psi_{k+1-p}) \Psi =0 \} \, .
	\end{equation}
	In other words, $F^p \Lambda^{p+q}$ is the space of forms of total degree $p+q$ and which have maximally $q$ factors in $\halg^\ast$. It is easy to check that $d$ maps
	$F^p \Lambda^r$ inside $F^{p} \Lambda^{r+1}$. Indeed, the degree is raised by one, and the number of factors in $\halg^\ast$ is raised by one maximally. 
	This implies that there exists a spectral sequence $(E_s,d_s)$ that converges to the Lie algebra cohomology $H^\ast(\galg)$ 
	\cite{BottTu}. We explain the part of this spectral sequence that we actually exploit. 
	In terms of the bidegree $(p,q)$ introduced above, the differentials $d_s$ have bidegree $(s,1-s)$ on the spaces $E_s$. For $E_0$ for instance, the operator $d_0$ increases the  $\halg$ index $q$ by one (and the total index by one, as always).
	The operator $d_1$ increases the  index $p$ by one (namely the $(\galg/\halg)^\ast$ index), and the second index $q$ by zero. 
	Let's make these spaces and differentials a little more concrete.
	
	In a first step, we introduce the (doubly graded) quotient spaces $E_0^{p,q}$
	\begin{equation}
	E_0^{p,q} = F^p \Lambda^{p+q} / F^{p+1} \Lambda^{p+q} \, .
	\end{equation}
	The elements of these spaces are represented by the forms that have exactly $q$ factors that lie in the subalgebra dual $\halg^\ast$. As a first approximation to the $d$ cohomology, we wish to compute the cohomology of the operator $d_0$ that $d$ induces on these spaces.  The operator $d_0$  is precisely the part of $d$ that augments the degree in $\mathfrak{h}^\ast$ by one (and the total degree by one, as always). A moment's thought shows that this differential $d_0$ acts on the forms in $\Lambda \mathfrak{h}^\ast$ like the ordinary Lie algebra differential (which in fact acts trivially on this
	abelian algebra) while it acts on $\Lambda (\galg/\halg)^\ast$ as if it were the action of the differential associated to the adjoint representation of $\halg$ on $\galg/\halg$, extended to the exterior product space. Thus, the action of $d_0$ is the action of the standard $\mathfrak{h}$ differential operator on differential forms that take values in a non-trivial representation of $\mathfrak{h}$  \cite{CE}.
	Therefore, the cohomology $E_1^{p,q}$ of the spaces $E_0^{p,q}$ with respect to $d_0$ coincides with  the Lie algebra cohomology for $\halg$ with values in the representation $\Lambda^p(\galg/\halg^\ast)$:
	\begin{equation}
	E_1^{p,q} \equiv H^q(\halg, \Lambda^p( \galg/\halg^\ast)) \, .
	\end{equation}
	We continue our approximation of the group cohomology by analyzing the differential $d_1$ induced by the differential $d$ on the intermediate cohomology spaces $E_1^{p,q}$. By general arguments \cite{BottTu}, the differential will
	augment the total degree by one and  the $\mathfrak{h}^\ast$ degree does not budge. A calculation shows that it is the relative Lie algebra homology differential (up to a sign) \cite{HochschildSerre}. The relative Lie algebra differential can be thought off as the Koszul differential $d$ that can be consistently restricted to the space of $\mathfrak{h}$ invariants in $\Lambda (\mathfrak{g}/\mathfrak{h})^\ast$. This is the second term term that we identified in the group cohomology differential $d$.  Since the 
	cohomology $E_2^{p,q}$ of the operator $d_1$ is trivial in the $\halg$ part, we obtain  the tensor product space \cite{HochschildSerre}:
	\begin{equation}
	E_2^{p,q}  \equiv H^q(\halg) \otimes H^p(\galg,\halg) \, , \label{E2}
	\end{equation}
	where  the second part follows from the definition of relative Lie algebra cohomology $H^p(\galg,\halg)$ \cite{CE,HochschildSerre}.
	
	We have arrived at the approximation of the Lie algebra cohomology that we were after. Our next step will be to understand the factor spaces on the right hand side of (\ref{E2}) and the action of the Weyl group on them. That will turn out to contain sufficient information to lift the approximation to an exact statement. 
	The ring $H^q(\halg)$ is trivial to compute since the differential is zero because the algebra $\halg$ is abelian. The cohomology corresponds to all differential forms on the dual of the Cartan subalgebra $\halg$, namely the exterior algebra $\Lambda \halg^\ast$. We need to determine the relative cohomology $H^p(\galg,\halg)$ as well as its properties under Weyl group transformations to make progress.
	That requires considerably more background, once again. On the one hand, one can compute the relative cohomology $H(\mathfrak{g},\mathfrak{h})$ by identifying it with the cohomology of the coset space $G/H$ where $G$ is a connected compact simple group with Lie algebra $\mathfrak{t}$ and $H$ is a maximal torus of the group $G$. We review how to compute this cohomology using Morse theory in appendix \ref{Coset}. Below, we follow a different  route that identifies the relative cohomology with a space of differential forms with harmonic polynomial coefficients.

	\subsubsection{Harmonic Polynomials}
	\label{HarmonicPolynomials}
	In this section we recall the definition and some properties of harmonic polynomials on the dual of a Cartan subalgebra $\mathfrak{h}$ of the simple Lie algebra $\mathfrak{g}$ \cite{Helgason}.
	Firstly, we need properties of the vector space $S \mathfrak{h}^\ast$ of polynomials on the dual $\mathfrak{h}^\ast$. There is an action of the Weyl group $W$ on the vector space $\mathfrak{h}^\ast$. The space $(S \mathfrak{h}^\ast)^W$of $W$ invariant polynomials in $S \mathfrak{h}^\ast$ is isomorphic to the space of $G$ invariant polynomials in $\mathfrak{g}^\ast$. This is Chevalley's restriction theorem.  Moreover, the space of such polynomials is generated by $\text{rank} \, \mathfrak{g}$ independent homogeneous polynomials $F_{i=1,\dots,\text{rank} \, \mathfrak{g}}$. The degrees of the polynomials are unique, and they multiply to give the order $|W|$ of the Weyl group $W$ \cite{Carter}. The Poincar\'e polynomial equals 
	\begin{equation}
P^{(S \mathfrak{g}^\ast)^{\mathfrak{g}} }= 	P^{(S \mathfrak{h}^\ast)^W} = \prod_{i=1}^{\text{rank} \, \mathfrak{g}} (1-t^{d_i})^{-1}
	\end{equation}
	where the $d_i$ are the degrees of the generators which are equal to the exponents $m_i$ of the simple Lie algebra plus one, $d_i=m_i+1$.
	
	We can identify the algebra of differential operators on $\mathfrak{h}$ with constant coefficients with the space of polynomials $S \mathfrak{h}$. 
	 The inner product of a derivation with a polynomial in $S \halg^\ast$ is the derivative of the polynomial evaluated at zero. This pairing is compatible with the group action since the inner product is.
	Finally, we can define the space of harmonic polynomials which is instrumental in our analysis of the group cohomology. The space of harmonic polynomials ${\cal H}$  is the space of polynomials which is annihilated by all strictly positive degree $W$-invariant differential operators in $S \halg$.
	
	We can now characterize the subspace of harmonic polynomials inside the full space of polynomials. In order to do that, we
define the ideal $J$ inside the space of polynomials $S \halg^\ast$ as the ideal   generated by the strictly positive degree $W$-invariant polynomials in $S \mathfrak{h}^\ast$. 
	The space of harmonic polynomials ${\cal H}$ is the orthogonal complement of the ideal $J$ inside the space of polynomials $S \mathfrak{h}^\ast$. 
	A consequence of this theorem is the expression for the Poincar\'e polynomial for the space ${\cal H}$ of harmonic polynomials:
	\begin{equation}
	P^{\cal H}(t) = P^{S \halg^\ast}/P^{(S \halg^\ast)^W } (t)= (1-t)^{-{\text{rank} \, \mathfrak{g}}} \prod_{i=1}^{\text{rank} \, \mathfrak{g}}(1-t^{d_i}) = \prod_{i=1}^{\text{rank} \, \mathfrak{g}}(1+t+\dots+t^{d_i-1}) \, . \label{PoincarePolynomialHarmonicPolynomials}
	\end{equation}
	Therefore, the dimension of the space of harmonic polynomials is $|W|$, the order of the Weyl group $W$.
	Moreover, the space of harmonic polynomials is isomorphic to the regular representation of the Weyl group \cite{Varadarajan}. We recall that the regular representation is a direct sum of irreducible representations with multiplicity equal to their dimension.
	
	Let us introduce the polynomial
	\begin{equation}
	\Pi = \prod_{\alpha \in \Delta^+} \alpha \, ,
	\end{equation}
	where $\Delta^+$ is the positive root system of the simple Lie algebra $\mathfrak{g}$.
	A useful fact is that the partial derivatives of $\Pi$ span the space of harmonic polynomials on $\halg^\ast$.
	Finally, we will need the following property of polynomials multiplied by differential forms. If we consider the harmonic polynomials tensored with the forms on $\halg$, then the algebra of Weyl group invariants is an exterior algebra with generators:
	\begin{eqnarray}
	d F_i' \, ,
	\end{eqnarray}
	where the $F_i$ are the invariant polynomials introduced previously, and 
	the prime indicates that we reduce the coefficients of the differential form modulo the ideal $J$ \cite{Reeder,Solleveld}.

	\subsubsection*{Example}
	We illustrate the generic statements above with an explicit example. For simplicity, we identify the algebras with their duals using the Killing form, and we work in terms of polynomials on the algebras. We consider once more the simple Lie algebra $\mathfrak{su}(3)$ and its Cartan subalgebra $\mathfrak{h} = \mathfrak{u}(1) \oplus \mathfrak{u}(1)$.	
	The invariant polynomials are the symmetric polynomials of order two and three:
	\begin{eqnarray}
	F_1 &=& 3 E_{\alpha_1} E_{-\alpha_1}+3 E_{\alpha_2} E_{-\alpha_2}+3 E_{\theta} E_{-\theta} +H_1^2 + H_1 H_2 + H_2^2
	\nonumber \\
	F_2 &=& 3 E_{\alpha_1} E_{\alpha_2} E_{-\theta} + 3 E_{-\alpha_1} E_{-\alpha_2} E_{\theta}
	-E_{\alpha_2} E_{\alpha_2} (2H_1+H_2)
	+ E_{\alpha_1} E_{-\alpha_1} (H_1+2 H_2)
	\nonumber \\
	& & 
	+ E_\theta E_{-\theta} (H_1-H_2)
	+ \frac{1}{9} (H_1-H_2) (2 H_1^2 + 5 H_1 H_2 + 2H_2^2) \, .
	\end{eqnarray}
	These group  invariant polynomials correspond uniquely to their Weyl invariant restrictions to the Cartan subalgebra $\mathfrak{h}$:
	\begin{eqnarray}
	F_1| &=& H_1^2 + H_1 H_2+ H_2^2
	\nonumber \\
	F_2| &=&  \frac{1}{9} (H_1-H_2) (2 H_1^2 + 5 H_1 H_2 + 2H_2^2) \, . \label{Weylinvariantpolynomials}
	\end{eqnarray}
	The Weyl group is generated by the two actions:
	\begin{eqnarray}
	H_1  \rightarrow  -H_1 & \qquad & H_2 \rightarrow H_1+H_2
	\nonumber \\
	H_1 \rightarrow H_1+H_2 & \qquad & H_2 \rightarrow -H_2 \, , \label{WeylGroupActionsu3}
	\end{eqnarray}
	which indeed leave the polynomials (\ref{Weylinvariantpolynomials}) invariant.
	The ideal $J$ generated by the projected invariant polynomials $F_1,F_2$ has a Gr\"obner basis 
	\begin{equation}
	H_1^2+H_1H_2+H_2^2, 2 H_1 H_2^2+H_2^3, H_2^4
	\, .
	\end{equation} 
	The vector space quotient of the polynomials by the ideal $J$
	is six dimensional 
and the harmonic polynomials give a basis which is orthogonal to the ideal $J$. The harmonic polynomials are given by
	\begin{equation}
	1, H_1,H_2, 2 H_1 H_2 + H_2^2, H_1^2 + 2 H_1 H_2 , H_1^2 H_2 + H_1 H_2^2 \, ,
	\end{equation}
	as can be checked by elementary differentiation using the Weyl invariant differential operators
	\begin{eqnarray}
	D_2 &=& \partial_{H_1}^2 - \partial_{H_1} \partial_{H_2} + \partial_{H_2}^2 \\
	D_3 &=&  \partial_{H_1}^2 \partial_{H_2}  - 
	 \partial_{H_1} \partial_{H_2}^2    \, ,
	\end{eqnarray}
	as well as using the orthogonality property. 
	 The harmonic polynomials indeed form a  regular representation of the Weyl group, as can be checked for instance  by direct action 
	 of the Weyl group on the Cartan subalgebra elements.	
	Moreover, we have that the polynomial $\Pi$
	\begin{eqnarray}
	\Pi &\propto & H_1^2 H_2 + H_1 H_2^2  = \text{highest degree harmonic} \, ,
	\end{eqnarray}
	and its partial derivatives indeed generate the space ${\cal H}$ of harmonic polynomials. 
	
	Finally, let us study the Weyl invariant differential forms with harmonic polynomial coefficients.
	Firstly, at each degree in $\Lambda \mathfrak{h}^\ast$, we have an irreducible representation of the Weyl group $W$. We have for $su(3)$ that they have dimensions $1,2 $ and $1$. They are respectively the trivial, the standard and the sign irreducible representations of $\mathfrak{su}(3)$. Therefore, there will be four Weyl group invariants.
Secondly, we compute the differentials $dF_i$ of the Weyl invariant polynomials:
	\begin{eqnarray}
	d F_1 &=& 
	 (2H_1 +H_2) dH_1 + (2H_2 + H_1) d H_2  = dF_1'
	\nonumber \\
	d F_2 &=& \frac{1}{3} \left( (2 H_1^2 + 2 H_1 H_2 -H_2^2) dH_1 + (H_1^2 - 2H_1 H_2 -2H_2^2) d H_2 \right) =dF_2' \, .
	\end{eqnarray}
	These differentials are already in reduced form, since the coefficients of the differentials are linear combinations of the harmonic polynomials.
	These are the two generators of the four-dimensional space of Weyl invariant differential forms with harmonic coefficients. Their Weyl invariance can once more be checked directly using the transformation rules (\ref{WeylGroupActionsu3}).

	\subsubsection{The Relative Cohomology}
	In the previous subsection, we acquainted ourselves with the harmonic polynomials. In this subsection, we discuss briefly the isomorphism between the relative cohomology $H^\ast(\mathfrak{g},\mathfrak{h})$ and the space of Weyl group invariants in the space of differential forms with harmonic polynomial coefficients.
To that end, we introduce the map $\psi$ that maps polynomials into differential forms of doubled degree:
	\begin{equation}
	\psi: S \mathfrak{g}^\ast \rightarrow \Lambda g^\ast \, .
	\end{equation}
	The  map is defined by  mapping:
	\begin{equation}
	\psi : \mathfrak{g}^\ast \rightarrow \Lambda^2 g^\ast: \xi \mapsto d \xi \, .
	\end{equation}
	Furthermore, the latter two-form is in the center of the exterior algebra because it is of even dimension. Therefore, we can extend $\psi$ to an algebra homomorphism
	$\psi: S \mathfrak{g}^\ast \rightarrow \Lambda \galg^\ast$. It is also a homomorphism of $G$ modules. This map can be restricted to the polynomials $S \mathfrak{h}^\ast$.  It then becomes 
	a map from the polynomials $S \halg^\ast$ into  the space of closed forms in $(\Lambda (\galg/\halg))^{\halg}$. It is also a homomorphism of $W$ modules.
	
	Moreover, we introduced the ideal  $J$ of $S \mathfrak{h}^\ast$ generated by the $W$-invariant polynomials of strictly positive degree. This ideal is a subset of the kernel of $\psi$, and we can divide the source by the ideal $J$. If we divide out by it in the source, and concentrate on the relative cohomology $H^\ast(\mathfrak{g},\mathfrak{h})$ in the target, then $\psi$ becomes an isomorphism $\bar{\psi}$ \cite{Borel,Reeder,Solleveld} from the harmonic polynomials onto the relative Lie algebra cohomology
	\begin{equation}
	\bar{\psi}: {\cal H}=S \mathfrak{h}^\ast/J \rightarrow H^\ast(\mathfrak{g},\mathfrak{h}) \, .
	\label{isomorphismpsibar}
	\end{equation}
%
%
%
%
	The Weyl group action on the harmonic polynomials on the left hand side is the regular representation, and the isomorphism, compatible with the Weyl group action,
	guarantees that the relative cohomology also carries the regular representation of the Weyl group. 

	Finally, we are ready to provide a more explicit description of the Lie algebra cohomology. We recall that on the second page of the Hochschild-Serre spectral sequence we found (\ref{E2}):
	\begin{equation}
E_2^{p,q} =	 H^p(\galg,\halg) \otimes  H^q(\halg) \, .
	\end{equation}
	Using the knowledge we gained on the Lie algebra cohomology, we can now write:
		\begin{equation}
	E_2^{p,q} =	 {\cal H}_2^p \otimes  \Lambda^q(\halg) \, .
	\end{equation}
	where ${\cal H}_2$ is the space of harmonic polynomials mapped into the relative cohomology via the map $\psi$, with the degree doubled (which we indicate with the lower index $2$ on ${\cal H}_2$). A crucial observation is now that the group cohomology $
	H^\ast(\mathfrak{g}) = (\Lambda \mathfrak{g}^\ast)^{\mathfrak{g}} $ corresponds to Weyl group invariants, and that the dimension of the space of Weyl group invariants inside the second page $E_2$ is equal to the dimension of $\Lambda^q(\halg)$. This is true because ${\cal H}_2$ is the regular representation of the Weyl group, which contains each irreducible representation of the Weyl group with multiplicity equal to its dimension. Thus, the spectral sequence projected onto Weyl invariants degenerates at the second page, and we obtain:
	\begin{equation}
	H^\ast(\mathfrak{g}) =
	 (H_2 \otimes \Lambda \halg^\ast)^W \, .
	\end{equation}
	Explicit representatives in this simplified picture of the Lie algebra cohomology are provided by the invariants discussed in subsection \ref{HarmonicPolynomials}. 
	We have that $(H \otimes \Lambda \mathfrak{h}^\ast)^W$ is  a free exterior algebra with the $\text{rank} \, \mathfrak{g}$ generators $dF_i'$ where these are the differentials computed from the invariant polynomials, with coefficients reduced to the space of harmonic polynomials.
	If we introduce a basis $\xi^i$ of the dual $\mathfrak{h}^\ast$ of the Cartan algebra, then an explicit set of $\text{rank} \, \mathfrak{g}$ generators for the algebra of gauge invariants is
	\begin{equation}
	\{ \sum_{i=1}^{\text{rank} \, \mathfrak{g}}\bar{\psi} (\partial_i {F}_j) \wedge \xi^i : 1 \le j \le \text{rank} \, \mathfrak{g} \} \, .
	\end{equation}
	Since the map $\bar{\psi}$ is degree doubling, we note that the total degree of generator $j$ equals $2m_j+1$. Thus, the Poincar\'e polynomial for the algebra of gauge invariants is
	\begin{equation}
	P_G(t) = \prod_{i=1}^{\text{rank} \, \mathfrak{g}} (1 + t^{2m_i+1}) \, ,
	\end{equation}
	as stated previously.

	\subsection*{The $\mathfrak{su}(3)$ Example}
	Let's illustrate  these theorems once more  in the  example of the Lie algebra $\mathfrak{g}=\mathfrak{su}(3)$. To compute the relative Lie algebra cohomology we compute
	in the space of $\mathfrak{h}$ invariants in $\Lambda (\mathfrak{g}/\mathfrak{h})^\ast$, spanned by the vectors:\footnote{We use again the isomorphism between the algebra and its dual provided by the Killing form.}
	\begin{eqnarray}
	1 ,
	\psi_{\alpha_1} \psi_{-\alpha_1}, 	\psi_{\alpha_2} \psi_{-\alpha_2}, 	\psi_{\theta} \psi_{-\theta}
	,
	\psi_{\alpha_1} \psi_{\alpha_2} \psi_{-\theta} , \psi_{-\alpha_1} \psi_{-\alpha_2} \psi_{\theta}
	& & \nonumber \\
	\psi_{\alpha_1} \psi_{-\alpha_1} 	\psi_{\alpha_2} \psi_{-\alpha_2}, 	\psi_{\alpha_2} \psi_{-\alpha_2} \psi_{\theta} \psi_{-\theta}, \psi_{\theta} \psi_{-\theta} 	\psi_{\alpha_1} \psi_{-\alpha_1} 
	,
	\psi_{\alpha_1} \psi_{-\alpha_1}	\psi_{\alpha_2} \psi_{-\alpha_2}	\psi_{\theta} \psi_{-\theta} \, .
	\end{eqnarray}
	We still need to compute the kernels and images of the Koszul differential $d$ in the space of $\halg$ invariants. 
	Using  the commutators amongst the non-Cartan generators, we find for instance that the closed cubic element is exact. The end result for the cohomology is:
	\begin{eqnarray}
	H^0 = 1 \, , \quad
	H^1 = \{ \} \, , \quad
	H^2 = \langle\psi_{\alpha_1} \psi_{-\alpha_1} +\psi_\theta \psi_{-\theta}, \psi_{\alpha_2} \psi_{-\alpha_2}+\psi_\theta \psi_{-\theta} \rangle  \, , \quad
	H^3 = \{ \}  \, ,  &&
	\nonumber \\
 H^5 =\{ \}  \, , \quad	H^4 = \langle    	\psi_{\alpha_2} \psi_{-\alpha_2} \psi_{\theta} \psi_{-\theta}, \psi_{\theta} \psi_{-\theta} 	\psi_{\alpha_1} \psi_{-\alpha_1}  \rangle  \, , \quad
	H^6 = \langle \psi_{\alpha_1} \psi_{-\alpha_1}	\psi_{\alpha_2} \psi_{-\alpha_2}	\psi_{\theta} \psi_{-\theta} \rangle \, . &&
	 \, .
	\end{eqnarray}
	The bijection between the harmonic polynomials and the relative cohomology proceeds by mapping
	the Cartan generators $H^i$ to the differentials as follows
	\begin{eqnarray}
	& & 
	-H^1 \leftrightarrow \psi_{\alpha_1} \psi_{-\alpha_1}+\psi_{\theta} \psi_{-\theta}  \, , \qquad -H^2 \leftrightarrow \psi_{\alpha_2} \psi_{-\alpha_2} +\psi_{\theta} \psi_{-\theta}\, , 
	\end{eqnarray}
	and similarly for higher degrees.
	We have the product of cohomologies:
	\begin{eqnarray}
	H(\halg) &=& \{ 1, \psi_1, \psi_2, \psi_{1} \wedge \psi_2 \}
	\nonumber \\
	H(\galg,\halg,F) &=& \{ 1, \psi_{\alpha_{1,2}} \psi_{-\alpha_{1,2}}+\psi_{\theta} \psi_{-\theta}, \psi_{\alpha_{1,2}} \psi_{-\alpha_{1,2}} \psi_{\theta} \psi_{-\theta}, 
	\psi_{\alpha_{1}} \psi_{-\alpha_{1}} \psi_{\alpha_{2}} \psi_{-\alpha_{2}} \psi_{\theta} \psi_{-\theta} \} \, .
	\end{eqnarray}
	The representation of the Weyl group on the Cartan subalgebra implies that $H(\halg)$ consists of a singlet, a standard doublet and a sign representation of the Weyl group $S_3$. The representation of the Weyl group on the relative cohomology is inherited from the Weyl group representation on the harmonic polynomials, which is the regular representation.  A Weyl group invariant is then found at level one, at level three, at level five and at level eight, as expected.\footnote{Note that these representatives are different from the invariant representatives described in subsection \ref{ExplicitInvariants}, but through the spectral sequence and ring isomorphisms, they span an equivalent ring.}
	
	In fact, from the spectral sequence analysis, one obtains a slightly more refined picture of the Lie algebra cohomology \cite{Reeder}. Since the exterior algebra $({\cal H} \otimes \Lambda \halg^\ast)^W$ has generators $dF_i' \in ({\cal H}^{m_i} \otimes \Lambda^1 \halg^\ast)$, we find the doubly graded multiplicity formula:
	\begin{equation}
	\sum_{n=0}^{(\text{dim} \, \galg - \text{rank} \, \galg)/2} \text{dim} Hom_W (\Lambda^q,{\cal H}^n) u^n = s_q (u^{m_1}, \dots , u^{m_{\text{rank} \, \galg}})
	\end{equation}
	with $s_q$ the elementary symmetric polynomial in ${\text{rank} \, \galg}$ variables.\footnote{We used that $m_1 + \dots + m_{\text{rank} \, \galg}=(\text{dim} \, \galg - \text{rank} \, \galg)/2$.}  Taking into account the degree doubling in the isomorphism $\bar{\psi}$ (\ref{isomorphismpsibar})  leads to the refined counting:
	\begin{equation}
	\sum_{i=0}^{\text{dim} \, \galg} \text{dim}( H^{i-q}(\galg,\halg) \otimes H^q(\halg))^W y^i =
	y^q s_q (y^{2m_1}, \dots, y^{2m_{\text{rank} \, \galg}}) \, .
	\end{equation}
	This formula codes how to the total degree of a Lie algebra cohomology element splits over the relative Lie algebra cohomology and the Cartan subalgebra cohomology.
	For instance, for the $\mathfrak{su}(3)$ algebra we find that for $q=0$, we have a single invariant which
	also lies at $i=0$. For $q=1$, we need to multiply $y(y^2+y^4)=y^3+y^5$ to find an invariant at total level $i$ equal to three and five, and for $q=2$ we compute $y^{2+2+4}=y^8$ to find the volume invariant. This is indeed what we saw in greater detail in our explicit calculations above.

	\subsubsection*{Summary}
	In this section, we  provided a  guide to the detailed mathematical literature that enables us to count and construct the eigenstates with zero eigenvalue in the supersymmetric quantum mechanics of section \ref{sqm}.

	\section{An Ideal Complement}
	\label{maximal}
	\label{subalgebra}
	In section \ref{GaugedLieAlgebraFermions} we concentrated on the zero eigenvalue subspace of the supersymmetric quantum mechanics model. It corresponds to the Lie algebra cohomology, or to the excitations in the theory with zero quadratic Casimir. These are the eigenspaces with minimal eigenvalue with respect to the positive definite Hamiltonian (\ref{HamiltonianCasimir}).
	In this section, we explore the eigenspaces of maximal eigenvalue at a given fermion degree $D$, and identify them with the exterior algebra $\Lambda \mathfrak{g}$  divided by the ideal $\langle Q \mathfrak{g} \rangle $ generated by the elements of the vector space $ Q \mathfrak{g}$. 
	The results we present are mostly from \cite{KostantEigenvalues,KostantOn}. We  streamline certain derivations, and obtain the results in terms of elementary calculations within our fermion supersymmetric quantum mechanics.

	\subsection{The Maximal Eigenvalue}
	To understand the space of maximal eigenvalue, it is advantageous to introduce the operator $\Box$  and study its properties:\footnote{Recall that we defined  the interior multiplication as the hermitian conjugate  of the exterior multiplication.}
	\begin{eqnarray}
	\Box &=& \epsilon( d \psi_a) i (d \psi^a) \label{BoxLimit1} \\
	&=& 	\frac{1}{4} f_{abc} \psi^b  \psi^c {f^a}_{de}  {\psi^\dagger}^d{\psi^\dagger}^e
	\nonumber \\
	&=& -\frac{1}{4} (f_{acd} {f^a}_{be} + f_{adb} {f^a}_{ce} ) \psi^b \psi^c  {\psi^\dagger}^d {\psi^\dagger}^e 
	\nonumber \\
	&=& \frac{1}{4} (f_{acd} {f^a}_{be} + f_{adb} {f^a}_{ce} ) \psi^b  {\psi^\dagger}^d  \psi^c {\psi^\dagger}^e 
	+ \frac{1}{4} \psi^b \psi^\dagger_b
	\nonumber \\
	& = &  -\frac{1}{2} \theta(\psi_a) \theta(\psi^a) + \frac{1}{2} \psi^b \psi^\dagger_b \nonumber 
	\\
	&=& \frac{1}{2} ( D-C_2(\theta)) \label{BoxLimit}
	\end{eqnarray}
	and a similar calculation holds for $\Box'$:
	\begin{equation}
	\Box' = i( d \psi_a) \epsilon (d \psi^a) = \frac{1}{2} ((\text{dim} \, \mathfrak{g} -D)-C_2(\theta)) \, . \label{BoxPrimeLimit}
	\end{equation}
	Our proof uses contractions and the Jacobi identity.\footnote{We avoid the use of  the Clifford algebra structure on $\Lambda \mathfrak{g}$, exploited in \cite{KostantOn}.}  {From} equation (\ref{BoxLimit}) and the positive definiteness of the operator $\Box$, we readily conclude that at given fermionic degree $D$, the maximal eigenvalue of the Hamiltonian $C_2(\theta)$ equals the degree $D$. (Similarly, it cannot be higher than $\text{dim} \, \mathfrak{g} - D$.) {From} now on, we concentrate on the maximal eigenvalue subspaces at given degree  $D$.

	\subsection{The Eigenspaces of Maximal Eigenvalue}
	We recall that the  eigenspaces are representations of the adjoint action, and as such decompose into irreducible highest weight representations. Moreover, we note that there is a weight basis of the exterior algebra given by decomposable elements.\footnote{A decomposable element of the exterior algebra is an element that can be written as a wedge product of elements in the underlying vector space.} The basis is given by the wedge product of weight vectors of the Lie algebra ${\mathfrak g}$. Thus, the highest weight states can be chosen to be decomposable, and we can compute the eigenvalue of the Hamiltonian directly on decomposable highest weight states \cite{KostantEigenvalues}.
	
	Consider a decomposable element $\psi$ of norm one. Consider the subspace $\mathfrak{a}$ of $\mathfrak{g}$ which is generated by the factors of the element $\psi$. Define a projection operator $P_{\mathfrak{a}}$ that projects an element of $\Lambda \mathfrak{g}$ orthogonally onto the subspace $\mathfrak{a}$. Define the operator $S_{\mathfrak a} = P_{\mathfrak a} H_{\mathfrak a} P_{\mathfrak a}$
	where $H_{\mathfrak a}$  is the Hamiltonian restricted to the subspace $\mathfrak a$. Define moreover the trace $s_{\mathfrak a}$ of the operator $S_{\mathfrak a}$. 
	In \cite{KostantEigenvalues} it is shown that for a decomposable element $\psi$ of norm one, we have that
	$\langle H\psi | \psi \rangle$ equals $k-s_{\mathfrak a}$ where $k$ is the degree $D$ of the decomposable element.
	Moreover, one can prove that the trace $s_{\mathfrak a}$ is zero if and only if $\mathfrak a$ is a commutative subalgebra of $ {\mathfrak g}$. Thus, maximal eigenvalue decomposable states $\psi$ correspond to commutative subalgebras $\mathfrak{a} \subset \mathfrak{g}$. 
	
	The detailed calculation is found in \cite{KostantEigenvalues}. We wish to provide some intuition. The Hamiltonian obtains a contribution of one from each factor in the decomposable element $\psi$, in a calculation reminiscent of the degree one calculation (\ref{degreeone}). However, if two factors do not commute, then there is a subtraction that arises from the non-zero commutator, as in the discussion of the degree two case, below equation (\ref{degreetwo}). That then spoils maximality of the eigenvalue \cite{KostantEigenvalues}.

	Using these descriptions of the maximal eigenvalue eigenspaces, one can  show that the eigenspaces $C^k$ of maximal eigenvalue equal to the degree	$k$
	correspond to the span of all elements that are wedge products of the basis elements of a $k$ dimensional commutative Lie subalgebra $\mathfrak{a}$ of $\mathfrak g$. Moreover, the total
	space $C=\sum_k C^k$ is the kernel of the operator $\Box$. Also, we have the orthogonal sum (with respect to the Killing form):
	\begin{equation}
	\Lambda \mathfrak{g} = C \oplus \langle Q \mathfrak{g} \rangle \, ,
	\end{equation}
	where $ \langle Q \mathfrak{g} \rangle $ is the ideal in $\Lambda g $ generated by the subspace $Q \mathfrak{g}$. This follows from the expression of the operator $\Box$ in terms of the exterior product $\epsilon(d \psi^a)$ (\ref{BoxLimit1}), and the nature of the inner product on $\Lambda \mathfrak{g}$ \cite{KostantOn}.

	\section{The Concrete Cohomology and Abelian Ideals}
	\label{explicit}
	\label{concrete}
	\label{abelianideals}
	In the previous section, we provided an abstract description of the spaces $C^k$ of maximal eigenvalue $k$ at fermion degree $k$ in terms of either the orthogonal complement of an ideal, or in terms of linear combinations of wedge products of commutative basis elements.
	In this section, we provide a more concrete description of the maximal eigenvalue space $C=\sum_k C^k$, and  the irreducible representations that appear at each fermion degree $k$. The description is based on  the  enumeration of the abelian ideals $\mathfrak{a}$ of a Borel subalgebra $\mathfrak b$ of the simple Lie algebra $\mathfrak g$. 
	
	\subsection{The Role of Abelian Ideals}
	Consider a commutative Lie subalgebra $\mathfrak{a}$ of $ {\mathfrak g}$. Define $[{\mathfrak a}]$  to be the wedge product of a basis of the subalgebra $\mathfrak{a}$. Then the maximal eigenvalue space 
	$C \subset \Lambda  {\mathfrak g}$ is the span of all $[{\mathfrak a}]$.  It is a graded submodule of $\Lambda  {\mathfrak g}$ with  respect to the adjoint action. Fix a Borel subalgebra $\mathfrak{b}$ of the Lie algebra $\mathfrak{g}$.
	Then one can show that for $\mathfrak{a}$ a commutative ideal of the Borel subalgebra $\mathfrak{b}$, the vector $[{\mathfrak a}]$ is a highest weight space in the space $C$ of maximal eigenvalues \cite{KostantEigenvalues,KostantOn}. It generates an irreducible representation of highest weight equal to  the sum $\Delta(\mathfrak{a})$ of the weights of the weight vectors in the abelian ideal $\mathfrak{a}$.
	Moreover, let $\Xi$ be an index set for all abelian ideals in the Borel subalgebra $\mathfrak{b}$. Let $C_\xi$ be the irreducible $\mathfrak{g}$ submodule of the exterior algebra $\Lambda  {\mathfrak g}$ generated by the highest weight vector $[\mathfrak{a}_\xi]$ where $\xi \in \Xi$.
	Then one  has the direct sum formula $C= \sum_{\xi \in \Xi} C_\xi$ \cite{KostantEigenvalues,KostantOn}. The multiplicities of highest weights are at most one, and the decomposition in irreducible representations is therefore unique. 
	
	We can approach the same classification problem from a slightly different angle. We wish to determine the decomposable highest weight states that correspond to highest weight states that factor into a basis of a commutative subalgebra. The quadratic Casimir Hamiltonian is easily evaluated on highest weight states with highest weight $\mu$ and equals
	\begin{eqnarray}
	H &=& (\rho+\mu)^2-\rho^2 \, ,
	\end{eqnarray}
	where $\rho$ is again half the sum of all positive roots of the simple Lie algebra $\mathfrak{g}$.
	A decomposable highest weight state with maximal eigenvalue therefore satisfies:
	\begin{equation}
	(\rho+\mu)^2-\rho^2 =k \, , \label{EqualityWeightDegree}
	\end{equation}
	where $k$ is the degree of the highest weight state.
	Again, one can  show that the  equality (\ref{EqualityWeightDegree}) holds if and only if the highest weight $\mu$ is the sum of $k$ positive distinct roots such that the corresponding root vectors span a commutative ideal of a Borel subalgebra $\mathfrak b$ of the Lie algebra $\mathfrak g$ \cite{KostantEigenvalues}.

	\subsection{The Abelian Ideals in a  Borel Subalgebra}
	We are still at a rather abstract level of the solution of characterizing the maximal eigenvalue eigenspaces. However,  a large number of properties of abelian ideals are known that makes their enumeration accessible \cite{Panyushev,Suter} . In this subsection, we sketch  these properties and illustrate them with several examples. We refer to \cite{Panyushev,Suter} for the proofs of the statements.
	
	There are $2^{\, \text{rank}  \, {\mathfrak g}}$ abelian ideals of a  Borel subalgebra $\mathfrak{b}$ of the simple Lie algebra $\mathfrak{g}$. The
	abelian ideals come in flags. Namely, if we denote the eigenspace of the algebra $\mathfrak{g}$ corresponding to the root $\beta$  by $\mathfrak{g}_{\beta}$, then abelian ideals  can be described as spaces
	$\mathfrak{g}_{\beta_1} \oplus \dots \oplus \mathfrak{g}_{\beta_k}$ where the degree $k$ can take any value $k=0,1,\dots,d$ where $d$ is the dimension of a maximal abelian ideal. This is because from an abelian ideal we can always strip off the space corresponding to a root with minimal length to obtain another abelian ideal. The set of ideals is therefore a partially ordered set, ranked by the dimension of the abelian ideals.
	
	An explicit description of abelian ideals is known.
	We define the subideal $\mathfrak{a}^{\not\perp \theta} \subset \mathfrak{a}$
	as the subspace of the ideal $\mathfrak{a}$ spanned by the root spaces corresponding to roots that are not orthogonal to the highest root $\theta$. The cardinality of the set of roots
	inside $\mathfrak{a}^{\not\perp \theta}$ is at most $h^\vee-1$ where $h^\vee$ is the dual Coxeter number of the simple Lie algebra $\mathfrak{g}$.
	Thus, all abelian ideals fall into sets characterized by their $\mathfrak{a}^{\not\perp \theta}$ subideal. Thus, we want to describe the sets of ideals $\mathfrak{a}^{\not\perp \theta}$
	and also the set of ideals $\mathfrak{a}$ for which this is the $\mathfrak{a}^{\not\perp \theta}$ subideal.
	
	The non-zero abelian ideals of the form $\mathfrak{a}^{\not\perp \theta}$ are in canonical 
	bijection with the set of long positive roots. 
	For each positive long root $\beta$, we define a length $L(\beta)=2 (\theta-\beta,\rho)/(\theta,\theta)$. For each positive root there is a unique Weyl element $w$ such that
	$w \beta=\theta$ and the length $l(w)$ of the Weyl element equals the length of the root $L(\beta)$.
	For such a Weyl group element $w$  we define the set of roots $ \Phi_w$ which equals the intersection of the set of positive roots $\Phi_+$ and the Weyl reflected set $w \Phi_-$ of negative roots, $\Phi_w=\Phi_+ \cap w \Phi_-$. We then have the non-zero abelian ideal with roots all non-orthogonal to the longest root: 
	\begin{equation}
	a^{\beta,min} = g_\theta \oplus \bigoplus_{\gamma \in \Phi_w} g_{\theta-\gamma} \, .
	\end{equation}
	There is a bijection between positive long roots and ideals $\mathfrak{a}^{\not \perp \theta}$ which maps $\beta$ to
	$\mathfrak{a}^{\beta,min}$.
	
	Thus, we have that each non-zero abelian ideal $\mathfrak{a}$ fits into a flag $\mathfrak{a}^{\beta,min} \subset \mathfrak{a} \subset \mathfrak{a}^{\beta,max}$ for some positive long root $\beta$ which is fixed by the equation $\mathfrak{a}^{\not\perp \theta}=\mathfrak{a}^{\beta,min}$. If the root $\beta$ is not perpendicular to the highest root then we have $\mathfrak{a}^{\beta,min} =\mathfrak{a} =\mathfrak{a}^{\beta,max}$. 
	Otherwise, we need to characterize the further ideals in the flag. To that end we define two Weyl groups.  For each positive long root $\beta$, we define the Weyl group $W_{\perp \beta}$  generated by  those Weyl reflections  $s_{\alpha_i}$ where the simple root $\alpha_i$ is perpendicular to $\beta$. Moreover, we define the subgroup of the affine Weyl group $\hat{W}_{\perp \beta}$ generated by simple reflections corresponding to simple roots orthogonal to $\beta$. Then the abelian ideals whose subideal is $\mathfrak{a}^{\beta,min}$ are in bijective correspondence with the right cosets $W_{\perp \beta} \setminus \hat{W}_{\perp \beta}$. Thus,  we have the one-to-one correspondence between abelian ideals and right cosets
	$\cup_{\beta \in \beta^{long}_+} \ W_{\perp \beta} \setminus \widehat{W}_{\perp \beta}$.
	One can be very explicit about on the one hand the Weyl group elements $w$ (see table pages 189-190 in \cite{Suter}) as well as the coset representatives (see table pages 197-201 in \cite{Suter}) for simple roots, to which all other cases reduce. These tables are sufficient to write down highest weights corresponding to  all the abelian ideals. Further properties of abelian ideals are known, but we restrict to providing explicit examples that partially  illustrate the general theory that we succinctly reviewed.
	
	\subsection{Examples}
	We study three examples, namely the algebra ${\mathfrak a}_1=\mathfrak{su}(2)$, the algebra ${\mathfrak a}_2=\mathfrak{su}(3)$ and the exceptional $\mathfrak{f}_4$ algebra.
	See also  \cite{Panyushev,Suter}.

	\subsubsection*{The Algebra $\mathfrak{a}_1$}
	The Lie algebra ${\mathfrak g}=\mathfrak{su}(2)$  has  a single positive (long and simple) root $\alpha$. There are two abelian ideals, which correspond to the empty set and the root $\alpha$.  Thus, the space $C$ consists of two terms: the trivial representation at degree zero and the adjoint representation at degree one. This agrees with the explicit calculation (\ref{characteristicpolynomialsu2}). It is indeed clear that the degree two and three elements are all elements of the ideal $\langle Q \mathfrak{g} \rangle$. To non-trivially illustrate the properties discussed above, we need more intricate examples.

	\subsubsection*{The Algebra $\mathfrak{a}_2$}
\begin{figure}[h]
		\begin{center}
			\begin{tikzcd}
			\{ \alpha_1+\alpha_2, \alpha_1 \} & & \{ \alpha_1 +\alpha_2, \alpha_2 \}  \\
			&  \{ \alpha_1+\alpha_2 \} \arrow[ur] \arrow[ul]  & \\
			& \{ \} \arrow[u] &  
			\end{tikzcd}
			\caption{Hasse Diagram of Abelian Ideals of the Borel Subalgebra of $\mathfrak{a}_2$ }
			\label{AbelianIdealsBorelA2Hasse}
		\end{center}
	\end{figure}
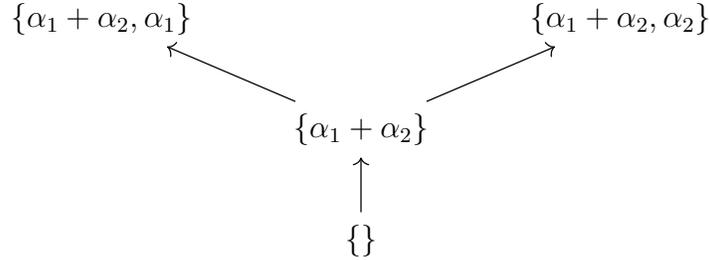

	The algebra ${\mathfrak g}=\mathfrak{su}(3)$ has dimension eight. The positive roots can be denoted $\alpha_1,\alpha_2$ and $\alpha_1+\alpha_2$. All roots are long and the roots $\alpha_{1,2}$ are simple. The abelian ideals are the vector spaces generated by the basis elements $\langle e_{\alpha_1+\alpha_2} \rangle, \langle e_{\alpha_1+\alpha_2},e_{\alpha_2} \rangle, \langle e_{\alpha_1+\alpha_2}, e_{\alpha_1} \rangle$ where $e_\beta$ is a Lie algebra generator with weight $\beta$. We conclude that the maximal eigenvalue space equals:
	\begin{equation}
	C = C^0 + C^1 + C^2 = 1 \oplus \xi_{\alpha_1+\alpha_2} \oplus (\xi_{2 \alpha_1 + \alpha_2} \oplus \xi_{ \alpha_1 + 2 \alpha_2} ) \, .
	\end{equation}
	If we check this with the list recorded in equation (\ref{characteristicpolynomialsu3}), we find a match using the relations between roots and fundamental weights, $3 \lambda_1 = 2 \alpha_1+\alpha_2$ and $3 \lambda_2 = \alpha_1+ 2 \alpha_2$.
	We  find four terms. There are no roots orthogonal to the highest root. Indeed, we have a bijection between the long positive roots and these ideals. We have the lengths $L(\alpha_i)=1$ and $L(\theta)=0$ and Weyl group elements $w_\theta=1,w_{\alpha_1}=s_{\alpha_2},w_{\alpha_2}=s_{\alpha_1}$ such that $\mathfrak{a}^{\theta,min}=\mathfrak{g}_\theta,
	\mathfrak{a}^{\alpha_1,min}=\mathfrak{g}_{\theta} \oplus \mathfrak{g}_{\alpha_1}$ and $ \mathfrak{a}^{\alpha_2,min}=\mathfrak{g}_{\theta} \oplus \mathfrak{g}_{\alpha_2}$. Thus, all of these ideals are minimal as well as maximal, in accordance with the general theory for abelian ideals corresponding to roots non-orthogonal to the highest root.
	%
	The Hasse diagram for these abelian ideals is given in figure \ref{AbelianIdealsBorelA2Hasse}.

	\subsubsection*{The Algebra $\mathfrak{f}_4$}
	For a more intricate illustration of the power of the general knowledge acquired on the abelian ideals in a Borel subalgebra of a simple Lie algebra, we turn to the exceptional Lie algebra $\mathfrak{f}_4$  \cite{Panyushev,Suter}. The  dimension of the algebra is fifty-two and therefore the dimension of the Hilbert space $\Lambda g$ is as large as $2^{52}$. The rank of the algebra is four and we  have twenty-four positive roots. These are given by the set:
	\begin{equation}
	\Delta_+ = \{ \epsilon_i, \epsilon_i\pm \epsilon_{j>i}, \frac{1}{2} (\epsilon_1 \pm \epsilon_2 \pm \epsilon_3 \pm \epsilon_4) \} \, ,
	\end{equation}
	where the quantities $\epsilon_{i \in \{ 1,2,3,4\} }$ form an orthonormal basis of a Euclidean vector space.
	The highest root $\theta$ equals $\theta=\epsilon_1+\epsilon_2=(2432)$. The latter notation gives the highest root as a positive integer linear combination of the simple roots  
	\begin{equation}\alpha_1 = \frac{1}{2}(\epsilon_1-\epsilon_2-\epsilon_3-\epsilon_4) \, ,
	\quad
	\alpha_2 = \epsilon_4  \, ,
	\quad 
	\alpha_3 = \epsilon_3-\epsilon_4 \, ,
	\quad 
	\alpha_4 = \epsilon_2-\epsilon_3 \, .
	\end{equation}
	Our conventions are those of \cite{OV,Panyushev}.
	There are twelve long positive roots and two simple long roots. We provide the Hasse diagram of the sixteen abelian ideals $I_n$ in figure \ref{AbelianIdealsBorelF4Hasse}. The set of positive roots corresponding to the building blocks  of the ideals is recorded in table \ref{AbelianIdealsBorelF4}.

	\begin{table}[h!]
		\centering
		\begin{tabular}{|c|c|c|}
			\hline 
			$I_n$ & $n$ & degree  \\
			\hline
			\{ \} & 0 & 0 \\
			\hline 
			$		\{ \theta \}$ & 1 & 1 
			\\
			\hline
			$ \{\theta,(2431) \}$ & 2 & 2 \\
			\hline
			$\{ \theta, (2431),(2421) \}$ & 3 & 3 \\
			\hline
			$\{ \theta, (2431),(2421),(2321) \}$ & 4 & 4 \\
			\hline
			$\{ \theta, (2431),(2421),(2321),(2221) \}$ & 5 & 5 \\
			\hline
			$\{ \theta, (2431),(2421),(2321),(1321) \}$ & 6 & 5 \\
			\hline
			$\{ \theta, (2431),(2421),(2321),(2221),(2211) \}$ & 7 & 6 \\
			\hline
			$\{ \theta, (2431),(2421),(2321),(1321),(2221) \}$ & 8 & 6 \\
			\hline
			$\{ \theta, (2431),(2421),(2321),(2221),(2211),(2210) \}$ & 9 & 7 \\
			\hline
			$\{ \theta, (2431),(2421),(2321),(2221),(2211),(1321) \}$ & 10 & 7 \\
			\hline
			$\{ \theta, (2431),(2421),(2321),(1321),(2221),(1221) \}$ & 11 & 7 \\
			\hline
			$\{ \theta, (2431),(2421),(2321),(2221),(2211),(2210),(1321) \}$ & 12 & 8 \\
			\hline
			$\{ \theta, (2431),(2421),(2321),(2221),(2211),(1321),(1221) \}$ & 13 & 8 \\
			\hline
			$\{ \theta, (2431),(2421),(2321),(1321),(2221),(1221),(0221) \}$ & 14 & 8 \\
			\hline
			$\{ \theta, (2431),(2421),(2321),(2221),(2211),(2210),(1321),(1221) \}$ & 15 & 9 \\
			\hline 
		\end{tabular}
		\caption{A Table of Abelian Ideals in a Borel Subalgebra of $F_4$ \cite{Panyushev,Suter}.}
		\label{AbelianIdealsBorelF4}
	\end{table}
	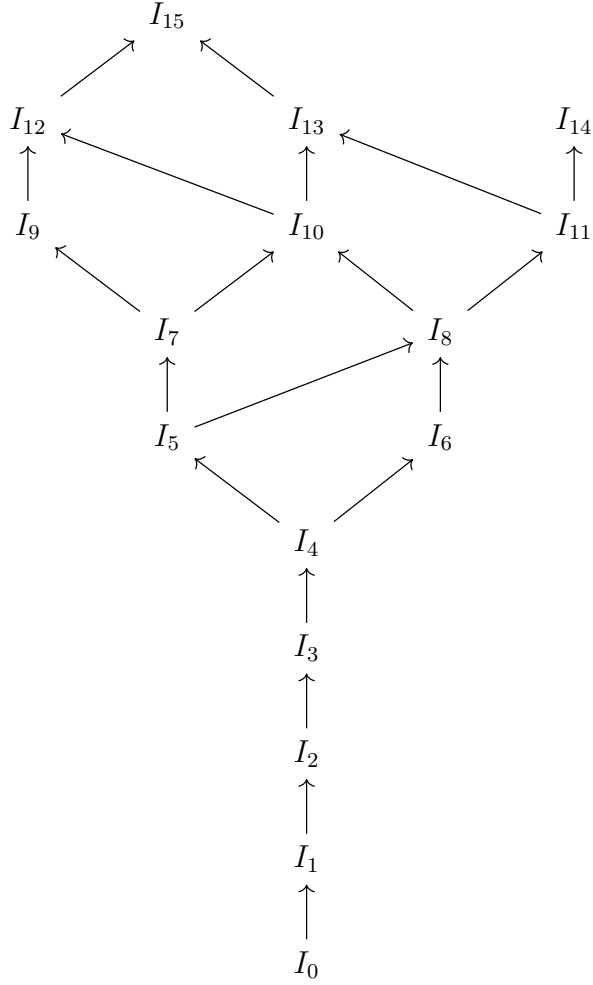
\begin{figure}[h]
		\begin{center}
			\begin{tikzcd}
			& I_{15} &    & & \\
			I_{12} \arrow[ur]	& & I_{13} \arrow[ul]  & & I_{14} \\
			I_9 \arrow[u] & & I_{10} \arrow[ull] \arrow[u] & &  \arrow[ull] \arrow[u] I_{11} \\
			&  I_7  \arrow[ul] \arrow[ur]&  & I_8  \arrow[ul] \arrow[ur] & \\
			&  I_5  \arrow[urr]  \arrow[u] &  & I_6 \arrow[u]& \\
			& & I_4  \arrow[ur] \arrow[ul] & & \\
			& &  I_3  \arrow[u] & & \\
			& & I_2\arrow[u] & & \\
			& & I_1\arrow[u] & & \\
			& & I_0 \arrow[u] & & 
			\end{tikzcd}
			\caption{Hasse Diagram of Abelian Ideals of the Borel Subalgebra of $\mathfrak{f}_4$ \cite{Panyushev,Suter}}
			\label{AbelianIdealsBorelF4Hasse}
		\end{center}
	\end{figure}
	
	In figure \ref{AbelianIdealsBorelF4Hasse}, the twelve long positive roots correspond bijectively to the ideals $I_{1-8},I_{10-11},$ $I_{13-14}$. The ideals $I_{9,12,15}$ are constructed from a subset of roots that contains a root orthogonal to the highest root $\theta=(2432)$, namely the root
	\begin{equation}
	(2210) = \epsilon_1 -\epsilon_2
	\, .
	\end{equation}
	The long simple roots $\alpha_{3,4}$ correspond to the ideals $I_{13,14}$ respectively. The length of the corresponding Weyl group element is  $h^\vee-2=7$.\footnote{The explicit expression for the Weyl group element of length seven can be found in \cite{Suter}.} 
	The Hasse diagram contains arrows that show the flag property of the Abelian ideals.
	
	Finally, the eigenspace of maximal eigenvalue at given fermion degree can be read off from table \ref{AbelianIdealsBorelF4}. The degree in the last column gives the degree which equals the eigenvalue. The highest weight of the irreducible representation is found by summing the entries characterizing the abelian ideals $I_n$ in the first column.
	To close, we remark that the number of abelian ideals of a given dimension for all simple Lie algebras $\mathfrak{g}$ was computed explicitly in \cite{Luo}. It answers the question how many irreducible representations make up the maximal eigenvalue eigenspace  at given degree.
	
	\subsubsection*{Summary and Sequel}
	Thus,  for the supersymmetric quantum mechanics model of section \ref{sqm}, we have described the zero eigenvalue subspaces in section \ref{GaugedLieAlgebraFermions}, and the maximal eigenvalue subspaces in sections \ref{maximal} and \ref{abelianideals}. As reviewed in subsection \ref{eigenvalues}, a few more results are known, but there is no closed solution to the problem of characterizing the representations at all eigenvalues.
	
	For the remainder of the paper, we will explore two tangential directions. As a preliminary, we note that
	the number of abelian ideals at degree below the dual Coxeter number $h^\vee$ has an interpretation in terms of the geometry of loop groups \cite{KostantPowers}. In fact, the theory of abelian ideals in Borel subalgebras is  related to affine Weyl group theory (as we already saw in the description of the ideals in terms of the affine Weyl group cosets). Thus, it becomes
natural to look for applications or relations of the previous results to fermions in one dimension higher. In section \ref{twodimensions}, we introduce one such model and show that it provides a cohomological interpretation of the maximal eigenvalue space $C$ \cite{KostantPowers}.  The two-dimensional model of massless fermions will  allow for the simplification of a key formula that is exploited in this interpretation.
	In section \ref{SYM}, we promote our fermions to four-dimensional fields and review how the supersymmetric quantum mechanics plays a role in the conjecture on the chiral ring of pure super Yang-Mills theory. We remark that one can identify canonical representatives of the chiral ring using the structures that we exhibted in the supersymmetric quantum mechanics.
	These sections illustrate the fact that the results on Lie algebra fermions summarized up to now have a wider applicability.
	
	\section{Lie Algebra Fermions in Two Dimensions}
	\label{twodimensions}
	We have studied a supersymmetric quantum mechanics model of Lie algebra valued fermions in $0+1$ dimensions in sections \ref{sqm} through \ref{abelianideals}.
	In this section we introduce Lie algebra valued fermions in $1+1$ dimensions. We have several motivations for increasing the dimension. The natural appearance of affine algebras in the mathematical analysis is one \cite{CP2,KostantPowers}. The second is that we will see that the maximal eigenvalue eigenspace $C$  corresponds to a cohomology once we study the model in two dimensions \cite{KostantPowers}.  More directly, our analysis teaches us properties of two-dimensional field theories of Lie algebra valued fermions. The most relevant mathematical references for this section are \cite{Garland,KumarBook,KostantPowers}.

	For starters, we define a $1+1$ dimensional field theory of massless Majorana fermions taking values in a simple Lie algebra $\mathfrak g$.\footnote{The zero modes  behave as the fermions in the model (\ref{CliffordAction}).} The massless fermions split into left-movers and right-movers and we concentrate on the left-movers only. We consider our model on a cylinder such that we can decompose the fermions in terms of Fourier modes along the spatial circle.
	We denote the positive frequency Fourier modes by $\psi_{-n <0}^a$ and the negative frequency modes by
	$\psi^{\dagger }_{m >0}$. The frequency is correlated with the mode number due to the massless and left-moving nature of the fermions. The dagger serves to indicate that we think of these operators as annihilation operators. Moreover, to remain close to the mathematical theory \cite{KostantPowers}, we will ignore the fermion zero modes initially.
	The modes of the real fermions satisfy the commutation relations:
	\begin{equation}
	\{ \psi_{-n}^a , \psi^\dagger_{b,m} \}  = \delta_{n,m} \delta_b^a \, . \label{affineanticommutator}
	\end{equation}
	The Hilbert space is created by the creation operators 
	\begin{equation}
	\psi_{-n\le -1}^a \, .
	\end{equation}
	It has a hermitian inner product. The Hilbert space can be thought off as the exterior algebra $\Lambda \mathfrak{u}^-$ of the subalgebra $\mathfrak{u}^-$ of the affine algebra $\hat{\mathfrak{g}}$ consisting of positive energy modes.
	We define a Hilbert space generated from a single vacuum state:
	\begin{eqnarray}
	\psi^{a_1}_{-n_1} \dots \psi^{a_k}_{-n_k} |0 \rangle \, .
	\end{eqnarray}
	As usual, after dimensional reduction on a circle, we can think of our  $1+1$ dimensional field theory as a supersymmetric quantum mechanics with an infinite number of degrees of freedom.

	\subsection{The Operators}
	\label{HamiltonianCalculation}
	As in the case of the supersymmetric quantum mechanics of section \ref{sqm}, we define interesting operators that serve as supercharges, and a Hamiltonian constructed out of the supercharges.  Our first main task will be to simplify the expression for the Hamiltonian. It will turn out that the Hamiltonian evaluated on an eigenstate equals the level of the state minus the quadratic Casimir.\footnote{The level of a state is given by the sum of the lower indices on the fermion  operators that create the state.}
	This property was first noted in \cite{Garland}\footnote{The calculation was omitted though the guiding remark "The collapsing is considerable." was offered.} and  a detailed published proof and generalisation
	can be found in \cite{KumarBook}. The proof of \cite{KumarBook} is clear. It requires considerable mathematical background. Rather than reviewing the proof in the mathematical language, we restrict to the affine Kac-Moody case we need here, and reprove the theorem using our fermion formalism. We present the details of the proof in appendix \ref{PhysicsKumarProof} where we comment on the relation between our calculation and the proof in \cite{KumarBook} without going through the lengthy translation from physics to mathematics in detail.\footnote{We have laid the groundwork for this doable and elaborate exercise in previous sections. We merely offer a few guiding remarks for readers familiar with the book \cite{KumarBook}.
		Let $ \hat{{\mathfrak g}}$ be the  (untwisted) affine Kac-Moody algebra corresponding to the simple Lie algebra $\mathfrak{g}$ in which the fermions take values. The algebra $\hat{\mathfrak{g}}$ takes the form $\hat{\mathfrak{g}}=\mathbb{C}[t,t^{-1}] \mathfrak{g} \oplus \mathbb{C} c \oplus \mathbb{C} d$ where $c$ is central and $d$ is the operator that measures the degree in the variable $t$. The degree in the auxiliary variable $t$ is physically interpreted as the oscillation or Fourier mode number $n$ which occurs as an index on our creation and annihilation operators. We write the Kac-Moody algebra as the direct sum
		$\hat{{\mathfrak g}}={\mathfrak n}^- \oplus \hat{{\mathfrak h}} \oplus {\mathfrak n}$ where the Cartan subalgebra is $\hat{\mathfrak{h}} = \mathfrak{h} \oplus  \mathbb{C} c \oplus \mathbb{C} d$ and $\mathfrak{h}$ is a Cartan subalgebra of $\mathfrak{g}$. The algebras $\mathfrak{n}$ and $\mathfrak{n}^-$ correspond to the positive and negative affine root spaces respectively. We have a standard maximal parabolic subalgebra ${\mathfrak p} = \sum_{n \ge0} {\mathfrak g} \otimes t^n \oplus \mathbb{C} c \oplus \mathbb{C} d $, and a nil-radical
		${\mathfrak u} =\sum_{n \ge 1} {\mathfrak g} \otimes t^n $.
		We also have that ${\mathfrak u}^-=\hat{\mathfrak{g}}/ \mathfrak{p}$. We note that there is  an adjoint action of the subalgebra $\mathfrak{p}$ on the vector space ${\mathfrak u}^-$.
		To be concrete, we can think of the generators $\psi_{-n\le -1}^a$ as a basis of the algebra $\mathfrak{n}^-$ and the generators $\psi^\dagger_{a,m \ge 0}$ can be thought off as corresponding to $ \sum_{n \ge0} {\mathfrak g} \otimes t^n$. The action of the central generator in the algebra $\mathfrak{p}$ is trivial while the operator $d$ acts to measure the degree of an operator. The other basis elements act as follows:
		\begin{equation}
		\bar{ad} (\psi^{\dagger b}_{m \ge 0}) ( \psi^a_{-n}) = {f^{ba}}_c \psi^c_{-n+m \ge -1} \, .
		\end{equation}
		The convention is that when the lower index on the right hand side violates the inequality, the expression is zero. See the book \cite{KumarBook}, page 98. This is a first and crucial step in translating the calculation in subsection 3.4 of \cite{KumarBook} into our fermionic language. See also appendix \ref{PhysicsKumarProof}.}
	
	Let us set up our infinite dimensional supersymmetric quantum mechanics.	
	Firstly, we define the  supersymmetry charges:
	\begin{eqnarray}
	\partial &=& \frac{1}{2} {f_a}^{bc} \psi^a_{-m-n} \psi^\dagger_{b,m} \psi^\dagger_{c,n}
	\nonumber \\
	\partial^\dagger &=&   \frac{1}{2} { f_{ab}}^c \psi^a_{-n} \psi^b_{-m} \psi^\dagger_{c,n+m}
	\end{eqnarray}
	where we understand all indices to be strictly of given sign (namely $-m-n \le -1$ et cetera). 
	We define a Hamiltonian $\Delta$ based on these supercharges: 
	\begin{eqnarray}
	\Delta  &=&  \partial \partial^\dagger +  \partial^\dagger \partial  \, . \label{AffineHamiltonian}
	\end{eqnarray}
	We also define the operator $L$ that measures the level of a state $\Psi$:
	\begin{equation}
	L \psi^{a_1}_{-n_1} \dots \psi^{a_k}_{-n_k} | 0 \rangle = (\sum_{s=1}^k {n_s}) \psi^{a_1}_{-n_1} \dots \psi^{a_k}_{-n_k} | 0 \rangle \, ,
	\end{equation}
	as well as the quadratic Casimir operator acting on the Hilbert space $\Lambda \mathfrak{u}^-$:
	\begin{equation}
	C_2 = f_{da b} \psi^a_{-n} \psi^{\dagger b}_n {f^d}_{e c} \psi^e_{-m} \psi^{\dagger c}_m \, .
	\end{equation}
	This is the quadratic Casimir for the adjoint action of $\mathfrak{g}$ on the Hilbert space. 
	Then, it is possible to prove (see \cite{KumarBook} and  appendix \ref{PhysicsKumarProof}) that when the Hamiltonian acts on a  state
	$\Psi$  of our Hilbert space $\Lambda \mathfrak{u}^-$ it simplifies to:
	\begin{equation}
	\Delta = \frac{1}{2}(L-C_2) = \frac{1}{2} (\text{Level} - \text{Quadratic Casimir})\, . \label{SimplifiedHamiltonian}
	\end{equation}
	Thus, if we have an irreducible $\mathfrak{g}$ representation component of the algebra $\Lambda \mathfrak{u}^-$ with highest weight $\mu$ and of level $N$,  the Hamiltonian in this representation equals:
	\begin{equation}
	\Delta = \frac{1}{2} ( N - (\rho+\mu,\rho+\mu)+(\rho,\rho)) \, ,
	\end{equation}
	where $\rho$ is half the sum of all positive roots for the algebra.
	This is a powerful result exploited in \cite{KostantPowers} to characterize the Lie algebra cohomology of the algebra $\mathfrak{u}^-$. Here, we remark on an alternative proof of the drastic simplification (\ref{SimplifiedHamiltonian}) of the Hamiltonian (\ref{AffineHamiltonian}).

	\subsection{An Alternative Proof}
	In this subsection, we provide an alternative manner (to appendix \ref{PhysicsKumarProof} and \cite{KumarBook}) to simplify the Hamiltonian (\ref{AffineHamiltonian}) of the $1+1$ dimensional field theory of left-moving Lie algebra fermions. The fermions are massless and the field theory exhibits conformal symmetry. We can harness the power of the infinite dimensional conformal symmetry in two dimensions to render the calculation easier.	Firstly, let's record the operator product expansion of the left-movers of our Lie algebra valued fields in the Neveu-Schwarz sector:\footnote{See e.g. \cite{Polchinski:1998rq,Polchinski:1998rr} for  background, conventions and nomenclature.}
	\begin{equation}
	\psi_a(z) \psi_b (w) =\frac{\kappa_{ab}}{z-w} + \text{reg} \, .
	\end{equation}
	The operator product expansion is consistent with the anti-commutation relation (\ref{affineanticommutator}), with shifted mode numbers.
	The operator product in the periodic Ramond sector has the same ultraviolet singularity.
	The purely fermionic theory under consideration exhibits an $N=1$ superconformal algebra with supercurrent $G$ and energy-momentum 
	tensor component $T$:
	\begin{eqnarray}
	G &=&  \frac{1}{3 \sqrt{2} } f^{abc} :\psi_a \psi_b \psi_c: \nonumber \\
	T &=& -\frac{1}{2} \kappa^{ab} : \psi_a \partial \psi_b :  \, ,
	\end{eqnarray}
	with central charge $c= \frac{1}{2} \, \text{dim} \,  \mathfrak{g}$. 
	We wish to exploit the known superconformal algebra operator product 
	\begin{equation}
	G(z) G(w) =  \frac{\frac{2c}{3}}{(z-w)^3} + \frac{T(w)}{z-w} + \text{regular} 
	\end{equation}
	which implies that the Fourier zero mode $G_0$ of the supercurrent $G$ in the Ramond sector and the Fourier zero mode $L_0$ of the energy-momentum tensor component $T$ satisfy
	\begin{equation}
	G_0^2 =  L_0 - \frac{c}{24}   =   \,\text{Level} \, .
	\end{equation}
	In the right hand side, we have registered the known fact that the zero mode of the energy-momentum tensor component measures the left-moving level, up to a normal ordering constant.
	With our goal in mind, we split the $N=1$ supercharge $G_0$  into terms that correspond respectively to the supercharges $\partial$, $\partial^\dagger$ as well
	as an extra term $E$ depending on the zero modes:
	\begin{eqnarray}
	G_0 &=& \frac{1}{\sqrt{2}} (f_{abc} \psi^a_{-n} \psi^b_{-m} \psi^{\dagger c}_{n+m} + f_{abc} \psi^a_{-n-m} \psi^{\dagger b}_n \psi^{\dagger c}_m)
	+ \sqrt{2} f_{abc} \psi^a_0  \psi^b_{-m} \psi^{\dagger c}_m + \frac{1}{3 \sqrt{2}} f_{abc} \psi^a_0 \psi^b_0 \psi^c_{0}
	\nonumber \\
	&=&  \sqrt{2} (\partial+ \partial^\dagger + E)
	\, ,
	\end{eqnarray}where
	\begin{eqnarray}
	E &=& f_{abc} \psi^a_0 \psi^{ b}_{-m} \psi^{\dagger c}_{m} + \frac{1}{6 } f_{abc} \psi^a_0 \psi^b_0 \psi^c_{0}  \, ,
	\end{eqnarray}
	and we have again used our index convention  that $\psi^a_{-n}$ has lower index less than or equal to minus one.
	We then  have the result:
	\begin{eqnarray}
	G_0^2 &=& 2(\partial + \partial^\dagger +E)^2 = L_0-\frac{c}{24}
	\nonumber \\
	&=& 2(\{ \partial , \partial^\dagger \} + E^2 + \{ \partial , E \} + \{  \partial^\dagger , E \} ) \, . \label{Intermediate}
	\label{prefinal}
	\end{eqnarray}
The operators $\partial$ and $\partial^\dagger$ square to zero. The anti-commutator of the operators $\partial$ and $\partial^\dagger$ equals the Hamiltonian we wish to compute.   Moreover, 
	the operators $\partial$ and $\partial^\dagger$ are invariant under the adjoint action. These operators also anti-commute with  the zero modes $\psi_0^a$. Therefore the latter two terms in the last line of (\ref{Intermediate}) are zero and we obtain:
	\begin{equation}
	2 \{ \partial , \partial^\dagger \} = L_0- E^2+\text{constant} \, .
	\end{equation}
	It remains to compute the part of the operator $E^2$ that does not depend on the zero modes. We must use the  anti-commutator of zero modes to obtain a non-zero contribution.
	We find in our Hilbert space $\Lambda \mathfrak{u}^-$:
	\begin{eqnarray}
	(E^2)_{\text{non-zero modes}} &=&  f_{abc} \psi^{\dagger b}_m \psi^c_{-m} {f^a}_{de} \psi^{\dagger d}_n \psi^e_{-n} + \text{constant}=   C_2 + \text{constant} \, .
	\end{eqnarray}
	We should note that we have  used two normal ordering schemes (namely singularity subtraction and oscillator normal ordering) that are inequivalent.   To fix the overall normal ordering constant proportional to the central charge it is sufficient to evaluate the left and right hand side of (\ref{prefinal}) on the  vacuum state.
	We conclude that we have the equality:
	\begin{equation}
	\{ \partial , \partial^\dagger \} =  \frac{1}{2} (\text{Level} -  \text{Quadratic Casimir} ) \, . \label{final}
	\end{equation}
	This is a short alternative derivation of the relation between the Hamiltonian, the level and the quadratic Casimir $C_2$. It would be interesting to investigate to what extent the vertex operator algebra derivation can be generalized to the broader Kac-Moody algebra context of \cite{KumarBook}.
	Given the brevity of our derivation, this could be worthwhile.

	\subsection{The Cohomology}
	\label{cohomology}
	Our second and final remark on the two-dimensional model at hand is that it provides  a  cohomological interpretation of the eigen\-spaces $C$ of maximal eigenvalue studied in sections \ref{maximal} and \ref{abelianideals}. This subsection reviews  part of \cite{KostantPowers}.
	The Hilbert space $\Lambda {\mathfrak u}^-$ is doubly graded by on the one hand the fermion number $k$ and on the other hand the level $N$, which is the absolute value of the sum of the lower indices of the creation operators $\psi^a_{-n}$. We have:
	\begin{equation}
	\Lambda {\mathfrak u}^- = \sum_{N \in \mathbb{N},n \in \mathbb{N}} (\Lambda^n {\mathfrak u}^-)_N \, .
	\end{equation}
	We introduce the grading $j$ which is the difference of the level and the fermionic degree, $j=N-n$. 
	The operator $\partial^\dagger$ raises the fermion degree by one and does not change the level. As such it acts as a map between the spaces
	$\Lambda^{[j]} \mathfrak{u}^- \rightarrow \Lambda^{[j-1]} \mathfrak{u}^-$ where the upper index in square brackets is the $j$-degree.
	Firstly, we observe what the states at $j$ degree zero look like.
	Since the fermion degree is equal to the level, and the level is at least one in the space $\mathfrak{u}^{-}$ for each fermion factor, we have that the degree zero subspace is built by the oscillators $\psi^a_{-1}$.
	Indeed, there is an isomorphism:
	\begin{equation}
	\Lambda {\mathfrak g} \rightarrow \Lambda^{[0]} {\mathfrak u}^-:
	\psi^1 \wedge \dots \wedge \psi^k \mapsto \psi^1_{-1} \wedge \dots \wedge \psi^k_{-1}
	\, . \label{isomorphism}
	\end{equation}
	Moreover, note that in the Hilbert space $\Lambda \mathfrak{u}^-$, we have that the spaces $\Lambda^{[j \le -1]} \mathfrak{u}^- =0$ are trivial.  Thus, the space $\Lambda^{[0]} \mathfrak{u}^-$ is automatically a subset of the kernel of the operator $\partial^\dagger$. Thus, to compute the $\partial^\dagger$ cohomology ${\cal H}^{[0]}$ at degree zero, we only need to determine which states at degree zero are exact. To that end, we start with a state at $j$-degree one, which must consist of (linear combinations of) states with all factors with index $-1$, and one factor with index
	$-2$. We then act with the supercharge $\partial^\dagger$ to generate the element:
	\begin{equation}
	\partial^\dagger \psi_{-2}^a \psi_{-1}^1 \dots \psi_{-1}^k | 0 \rangle =
	\frac{1}{2} {f^a}_{bc} \psi_{-1}^b \psi_{-1}^c \psi_{-1}^1 \dots \psi_{-1}^k | 0 \rangle \, . \label{Lambda1action}
	\end{equation}
	We note that the operation $\partial^\dagger$ acting on the space $\Lambda^{[1]} \mathfrak{u}^-$ as in equation (\ref{Lambda1action}) after applying the isomorphism (\ref{isomorphism}) precisely generates the  ideal $\langle Q \mathfrak{g} \rangle$ inside the exterior algebra $\Lambda \mathfrak{g}$. Thus we have that \cite{KostantPowers}
	\begin{equation}
	{\cal H}^{[0]} = \Lambda^{[0]} \mathfrak{u}^- / \partial^\dagger ( \Lambda^{[1]} \mathfrak{u}^-)
	=   \Lambda \mathfrak{g} / \langle d \mathfrak{g} \rangle = C \, . 
	\end{equation}
	This provides a cohomological interpretation to the quotient space $C$ of maximal eigenvalue, studied in sections \ref{maximal} and \ref{abelianideals}.

	\section{The Chiral Ring of Pure Super Yang-Mills}
	\label{SYM}
	\label{chiralring}
	
	In this second tangential section, we recall the conjectured structure of the chiral ring of pure supersymmetric Yang-Mills theory in four space-time dimensions \cite{Cachazo:2002ry}, and the status of its proof. We then offer a few remarks that may be of use in the construction of a uniform proof of the conjecture. As a by-product, we  argue that there is a canonical representation of the chiral ring. We obtain the canonical representation in the example where the gauge algebra $\mathfrak{g}$ is $\mathfrak{su}(3)$. 
	\subsection{The Context and the Conjecture}

	The supersymmetric quantum mechanics problem discussed in this paper has applications in supersymmetric gauge theory. Indeed, some of the results presented here in the context of supersymmetric quantum mechanics formed the basis for an attempt to prove a conjecture in pure supersymmetric Yang-Mills theory. The conjecture in pure supersymmetric Yang-Mills theory with simple
	gauge algebra $\mathfrak{g}$ is as follows \cite{Cachazo:2002ry}: the classical chiral ring of pure super Yang-Mills theory is generated by a single generator $S$, the glueball field, that satisfies $S^{h^\vee}=0$ as well as  $S^{h^\vee-1} \neq 0$. 
	Let us present the ingredients in the conjecture in more detail, and briefly recall what has been proven.
	
	Firstly, we recall that all chiral operators in pure supersymmetric Yang-Mills theory are functions of  the vector multiplet superfield $W_\alpha$. Chiral operators are defined by the fact that they
	are  annihilated by a set of supercharges $\bar{Q}_{\dot{\alpha}}$. Secondly, the operators $[ \bar{Q}_{\dot{\alpha}},O]$ are by definition $\bar{Q}_{\dot{\alpha}}$  exact. If we concentrate on correlators of chiral operators, then exact operators decouple from those correlators.  Thus, the chiral correlation functions only depend on the $\bar{Q}_{\dot{\alpha}}$ cohomology. Finally, we recall the fact that the commutator of vector multiplet superfields, 
	$[W_\alpha,W_\beta]$, is $\bar{Q}_{\dot{\alpha}}$ exact. Importantly, the operator that is the pre-image of the commutator is not itself chiral. 
	In summary, there is an interesting cohomological problem to solve in four dimensions which is to compute the $\bar{Q}_{\dot{\alpha}}$ cohomology and its ring structure.\footnote{To tie in to earlier sections we note  that when we reduce the four-dimensional cohomological problem to the supersymmetric quantum mechanics of Lie algebra fermions corresponding to $W_1$ and $W_2$, than the problem is no longer cohomological (since the pre-image of the commutator is not part of  the supersymmetric quantum mechanics). A consequence of the  remark in subsection \ref{cohomology} is that it is sufficient to study a two-dimensional problem to retain a cohomological interpretation.}
	
	Let us  review the status of the conjectured solution recalled above. The conjecture has been proven for all the classical groups \cite{Cachazo:2002ry,Witten:2003ye}, at first on a case-by-case basis, and later uniformly \cite{Etingof:2003fy}. Moreover, it was proven for the exceptional Lie algebra $\mathfrak{g}_2$ \cite{Etingof:2003dd}. For all cases, it is known that the glueball operator $S=\text{Tr} W_\alpha W^\alpha = 2 W_1^a W_{2a}$ generates the ring.
	Moreover, there are conjectures that in turn imply the conjecture, which chop up the general problem into potentially more tractable ones \cite{Kumar2,Etingof:2003fy}. However, there is no proof of the conjecture for general Lie algebra $\mathfrak{g}$, let alone a uniform proof. Below we provide minor extra observations on the state of the problem.

	\subsection{Structural Remarks}
	Firstly, note with \cite{Witten:2003ye,Etingof:2003dd,Kumar2} that for one component of the superfield $W_\alpha$, the problem of computing the cohomology (without demanding gauge invariance at first) is equivalent to computing the space $C$ which is orthogonal to the ideal $\langle d \mathfrak{g} \rangle$ generated by the commutator inside the exterior algebra $\Lambda \mathfrak{g}$. See sections \ref{maximal} and \ref{abelianideals}.
	We note that the space $C$  is in fact canonically defined as the orthogonal complement to the ideal. Thus, we obtain {\em canonical} representatives for the quotient space. 
	
	However, we have two components to the superfield $W_\alpha$, and we therefore work with the direct sum algebra $\mathfrak{g} \oplus \mathfrak{g}$ as well as the exterior algebra $\Lambda \mathfrak{g} \otimes \Lambda \mathfrak{g}$. Still, the same remark applies to the second component of the superfield $W_\alpha$. When we project onto gauge invariants (namely invariants under the diagonal Lie subalgebra that acts in the adjoint), then we obtain the canonical diagonal invariants inside $C_1 \otimes C_2$  inside the exterior algebra $\Lambda \mathfrak{g} \otimes \Lambda \mathfrak{g}$ as described in \cite{Etingof:2003dd}. There is a single invariant for each irreducible representation in $C$ since the irreducible representations all have multiplicity one. The remaining task is to mod out the  space of invariants with the commutators of the components $W_1$ with $W_2$. Recall that we have an explicit description of the spaces $C_i$ in terms of irreducible representations and that we therefore have a hands-on description of the full space of invariants. In particular, for instance, we know the dimension of the space of invariants explicitly at each degree.
	
	When we study this problem in the light of how we obtained the spaces $C_{1,2}$ in the first place, we notice that it is natural to again attempt to compute the orthogonal complement to the ideal
	generated by the commutators $[ W_1, W_2]$. Again, we can compute the orthogonal complement as the space annihilated by the operator conjugate to the exterior product, namely the operator
	$i_{12}^a = {f^a}_{bc} \psi_2^{\dagger b} \psi_1^{\dagger c}$. The lower indices refer to the two copies of the supersymmetric quantum mechanics model  introduced in section \ref{sqm} that are necessary to model the two components of the superfield $W_{\alpha=1,2}$. Thus, we can mimic for instance our analysis in section \ref{sqm} and see whether the operator allows for 
	a major simplification, e.g. can be summarized in a BPS equation of the type found in \cite{KostantEigenvalues}.\footnote{See equation (\ref{EqualityWeightDegree}).} However, this is optimistic. Indeed, the symmetry properties of the problem are considerably less constraining than those governing the spectrum of the Laplacian on the single copy of the exterior algebra. This is because the symmetry of the problem is only the diagonal Lie subalgebra, and we have already projected onto invariants of the diagonal subalgebra. Therefore, the analysis of this operator and its kernel will be of a different type. 
	
	We proceed a little further with this approach and characterize the resulting problems in  more detail. We note that the operator $\Box_{12} = i_{12a}^\dagger i_{12}^a$ is a scalar operator that respects both the degree operator $D_1$ with respect to the first exterior algebra, as well as the degree operator $D_2$ with respect to the second.
	We can then attempt to analyze the operator $\Box_{12}$ as we did previously (see e.g. equation (\ref{BoxLimit})) and manipulate:
	\begin{eqnarray}
	\Box_{12} &=&
	{f}_{abc} \psi_1^{b} \psi_2^{c}  {f^a}_{de} \psi_2^{\dagger d} \psi_1^{\dagger e}  \nonumber \\
	&=&  - (f_{acd} {f^a}_{be} + f_{adb} {f^a}_{ce})  \psi_1^{b} \psi_2^{c}\psi_2^{\dagger d} \psi_1^{\dagger e} 
	\nonumber \\
	&=& 	 (f_{acd} {f^a}_{be} + f_{adb} {f^a}_{ce})  \psi_1^{b} \psi_2^{\dagger d} \psi_2^{c} \psi_1^{\dagger e} 
	+ \psi_1^{b}  \psi_{1b}^{\dagger } 
	\nonumber \\
	&=&  -f_{abd} {f^a}_{ce}  \psi_1^{b} \psi_2^{\dagger d} \psi_2^{c} \psi_1^{\dagger e} 
	-f_{acd} {f^a}_{be} \psi_2^{c} \psi_2^{\dagger d}   \psi_1^{b} \psi_1^{\dagger e}
	+ \psi_1^{b}  \psi_{1b}^{\dagger }  \nonumber \\
	&=& -f_{abd} {f^a}_{ce}  \psi_1^{b} \psi_2^{\dagger d} \psi_2^{c} \psi_1^{\dagger e}  - \theta_2 (\psi_2^a) \theta_1(\psi_{1a}) + D_1  \, .
	\end{eqnarray}
	Because the operator is degree compatible, we can work at fixed degree and put
	$D_i=k_i$. Also, we can restrict the analysis to diagonal $\mathfrak{g}$ invariants and the adjoint actions $\theta_1$ and $\theta_2$ with respect to $\mathfrak{g}_{1,2}$ then combine into a trivial action. We thus have
	$(\theta_1+\theta_2)^2=0$ and since $\theta_i^2=k_i$ on the invariants orthogonal to $\langle dg_i \rangle$, we can simplify this equation to  $-2 \theta_1 \theta_2 = k_1+k_2$. 
	We thus find:
	\begin{eqnarray}
	\Box_{12} &=& 
	-f_{abd} {f^a}_{ce}  \psi_1^{b} \psi_2^{\dagger d} \psi_2^{c} \psi_1^{\dagger e}  + \frac{k_1+k_2}{2} + k_1 
\, .
	\end{eqnarray}
	We can moreover exchange $1,2$ in the equation, and recombine:
	\begin{eqnarray}
	\Box_{12}&=&
	-f_{abd} {f^a}_{ce}   \psi_2^{c} \psi_1^{\dagger e}  \psi_1^{b} \psi_2^{\dagger d}+
	\frac{k_1+k_2}{2} + k_2 
	\nonumber \\
	&=& -\frac{1}{2} f_{abd} {f^a}_{ce}   (\psi_2^{c} \psi_1^{\dagger e}  \psi_1^{b} \psi_2^{\dagger d}+\psi_1^{c} \psi_2^{\dagger e}  \psi_2^{b} \psi_1^{\dagger d}
	) + k_1+k_2
	\, . \label{Reformulation}
	\end{eqnarray}
Unfortunately, we have found the quartic fermion operator that makes up the first term in the right hand side of equation (\ref{Reformulation}) hard to evaluate. We have reformulated one difficult problem in terms of a different one.

	Let us make a few more obervations.
	Firstly, from any diagonal $\mathfrak{g}$ invariant, the operator $i_{12}^a$ either generates an adjoint representation, or zero. In other words, the result of the action of $i_{12}^a$ is either zero or non-zero for all indices $a$ simultaneously. Thus, we can analyze whether this operator is zero for any given index $a$, and then it will be true for all indices $a$. A preferred choice would be to take $a$ equal to the highest root $\theta$.
	An important observation is likely that abelian ideals of the Borel subalgebra that only have roots that are not perpendicular to the highest root $\theta$ are of maximal dimension $h^\vee-1$.
	The associated Heisenberg algebras are described in \cite{KostantPowers}. Moreover, it is known that there exists a relation between invariants at degree $h^\vee$ \cite{CPP,Etingof:2003fy}. We speculate that a proper combination of these observations may help in building a uniform proof of the conjecture. Finally, we note that the number of abelian ideals at given degree grows or remains constant 
	until we reach the dual Coxeter number, after which it strictly decreases \cite{Luo}. This observation may be helpful in proving that  the operator $\Box_{12}$ admits precisely one zero eigenvalue in the space of invariants of degree below  the dual Coxeter number, namely, it is of maximal rank minus one.
	In the next section, we illustrate some of these remarks in the concrete example of $\mathfrak{g}=\mathfrak{su}(3)$.

	\subsection{The Example}
	\label{SYMExample}
	
	We recall that in the case of the Lie algebra $\mathfrak{su}(3)$, the space $C$ of maximal eigenvalues consists of the invariant at level zero,  an adjoint at level one and a $10$ and $\bar{10}$ representation at level two. Thus, in the tensor product space $C_1 \otimes C_2$, we have one invariant at level zero (with respect to fermionic degree $D_1$ as well as with respect to fermionic degree $D_2$), one at level one and two at level two. The invariants at level zero and one obviously obey the third constraint,  namely that $i_{12}^a$ acting on them equals zero. There are two invariants at level two and we want to understand how a single linear combination survives the third constraint. To that end, we can explicitly construct the two invariants and then act with the operators $i_{12}^a$ on a generic linear combination, and solve for the zero modes in terms of the two coefficients of the invariants.  A single linear combination must survive at level two. Generically, we must find exactly one solution to the equations, which therefore must have maximal rank minus one. 
	
	The invariants in $C_1 \otimes C_2$ for the Lie algebra $\mathfrak{g}=\mathfrak{su}(3)$ at level two are made from the maximal eigenvalue eigenspaces which form the irreducible representations $10$ and $10'$ of $\mathfrak{su}(3)$. The weights of these representations have multiplicity equal to one, and they are given by:
	\begin{eqnarray}
	10 &:&  2 \alpha_1+\alpha_2, \alpha_1+\alpha_2,\alpha_1,\alpha_2, -\alpha_1+\alpha_2, 0, - \alpha_1, -\alpha_2, -\alpha_1-\alpha_2,-\alpha_1-2 \alpha_2
	\nonumber \\
	\bar{10} &:&  2 \alpha_2+\alpha_1, \alpha_1+\alpha_2,\alpha_1,\alpha_2, -\alpha_2+\alpha_1, 0, - \alpha_1, -\alpha_2, -\alpha_1-\alpha_2,-\alpha_2-2 \alpha_1 \, .
	\end{eqnarray}
	We explicitly computed the states at level two with these weights.\footnote{ See Appendix \ref{su3canonicalrepresentatives} for a few details.} We normalized the resulting invariants $I_{10}$ and $I_{10'}$  such that one obtains one from the other by exchanging the two simple roots $\alpha_{1,2}$. We acted on them with the  operators $i_{12}^a$. We found that the result vanishes if and only if the coefficients of the two invariants are equal. Thus, the canonical representative of the pure super Yang-Mills chiral ring for $\mathfrak{g}=\mathfrak{su}(3)$ at fermion number two equals $I_{10}+I_{10'}$.
	The point of the example is to illustrate the general logic as well as the fact that at the end of the procedure one obtains canonical representatives of the chiral ring.

	\section{Conclusions}

	\label{conclusions} 
	
	We studied the quantum mechanics of simple Lie algebra valued fermions. There are natural supercharges and Hamiltonian operators that act in the fermionic Hilbert space, and they correspond to standard operations in the mathematics literature on the exterior algebra $\Lambda \mathfrak{g}$ of the simple Lie algebra $\mathfrak{g}$. We translated the mathematics theorems into  physical  properties of Lie algebra valued fermions, and in a few cases, simplified steps in proofs of  properties of the spectrum of the  fermion model. 
	We identified the eigenspace of zero eigenvalue with the Lie algebra cohomology, and reviewed how to compute the latter. We also summarised what is known about the eigenspace of maximal eigenvalue at given fermion degree. We embedded the cohomological interpretation of the quotient of the exterior algebra by the ideal generated by the supercharge derivative operator in a two dimensional model of Lie algebra fermions. The embedding provided an opportunity to simplify a lengthy mathematical proof using operator product expansions in conformal field theory. We also recalled the relevance of these models for the conjecture on the chiral ring of pure supersymmetric Yang-Mills theory in four dimensions, and explained how to determine canonical representatives of this ring. 
	We hope our efforts at simplification and translation will make the attack on a universal proof of the conjecture for any simple Lie algebra $\mathfrak{g}$ successful. 
	
The powerful mathematical results we reviewed are of relevance to a large number of field theories with Lie algebra valued fermions. We have rendered these theorems more accessible to physicists and are confident that this will give rise to further cross fertilisation between  these domains.

	\section*{Acknowledgments}
	It is a pleasure to thank my colleagues  for creating a stimulating research environment.

	\appendix

	\section{The Killing Form}
	\label{Killing}
	
	The Killing form $(,)$ is equal to the trace in the adjoint representation:
	\begin{equation}
	(\psi,\chi) = \text{tr} (\text{ad}(\psi) \text{ad}(\chi))
	\end{equation}
	where $\text{ad}(\psi)(\phi)=[\psi,\phi]$ is the adjoint action on an element $\phi$ of the simple Lie algebra $ {\mathfrak g}$. 
	If we pick a basis $\psi^a$ such that $[ \psi^a, \psi^b ] = {f^{ab}}_c \psi^c$, then the Killing form evaluates to
	\begin{equation}
	(\psi^a,\psi^b)_\kappa = {f^{ac}}_d {f^{bd}}_c = \kappa^{ab} \, ,
	\end{equation}
	where we introduce the symbol $\kappa^{ab}$ for the inverse Killing metric. The Killing metric is positive definite on
	$\mathfrak{q}=i \mathfrak{t}$ where $\mathfrak{t}$ is the Lie algebra of a compact group associated to the simple Lie algebra $\mathfrak{g}$.

		 \section{Low Rank Algebras: Conventions and Calculations}
		 \label{ExplicitAlgebras}
		 We record a few of our conventions and technical results. 
		 \subsection{The Lie Algebra $\mathfrak{su}(2)$}
The Lie algebra $\mathfrak{su}(2)$ has a basis $\{ H, E_{\alpha}, E_{-\alpha} \}$ with commutation relations:
\begin{equation}
[H,E_{\pm \alpha} ] = \pm 2 E_{\pm \alpha} \, , \qquad [E_\alpha, E_{-\alpha}] = H \, .
\end{equation} 
The Killing metric in this space is 
\begin{equation}
\kappa_{ab} = 
\left( \begin{array}{ccc}
 2 & 0 & 0 \\
0 & 0 & 1 \\
0 & 1 & 0 
        \end{array} \right) \, .
\end{equation}
		 We use the Killing metric to raise and lower indices on the Lie algebra valued fermions $\psi_H, \psi_{\pm \alpha}$ that correspond to the generators $H,E_{\pm \alpha}$ respectively under the identification $\Lambda^1 \mathfrak{g}=\mathfrak{g}$.\footnote{We note that the Lie algebra basis for the complexified Lie algebra that we choose in this appendix differs from the real basis that is handy for the abstract calculations in the bulk of the paper. }
		 
		 \subsection{The Lie Algebra $\mathfrak{su}(3)$}
	The Lie algebra $\mathfrak{su}(3)$ has a basis $\{ H_1, H_2, E_{\pm \alpha_1}, E_{\pm \alpha_2}, E_{\pm \theta} \}$. The  positive simple roots are $\alpha_{1,2}$ and the third and highest positive root is $\theta=\alpha_1+\alpha_2$.
		 The generators satisfy the commutation relations, among others:
		 \begin{eqnarray}
 [H_1, E_{\pm \alpha_1}] = \pm 2 E_{\pm \alpha_1}	\, \qquad [H_1,E_{\pm \alpha_2}] = \mp E_{\alpha_2}	 \, \qquad [H_1, E_{\pm \theta}] = \pm E_\theta
  &&
\nonumber \\
{[} E_{\alpha_{1,2}}, E_{- \alpha_{1,2}} ] = H_{1,2}   \, \qquad [E_{\alpha_1},E_{\alpha_2}] = E_{\theta} \, \qquad [ E_{\theta}, E_{-\theta} ] = H_1 + H_2 \, . &&
		 \end{eqnarray}
		 Again these generators correspond straightforwardly to the fermions $\psi_{1,2},\psi_{\pm \alpha_{1,2}}$ and $\psi_{\pm \theta}$.
		 The Killing metric is
		 \begin{equation}
		 \kappa_{ab} = \left( 
		 \begin{array}{cccccccc}
		 	2 & -1 & 0 & 0 & 0 & 0 & 0 & 0 \\
		 	-1 & 2 & 0 & 0 & 0 & 0 & 0 & 0 \\
		 	0 & 0 & 0 & 0 & 1 & 0 & 0 & 0 \\
		 	0 & 0 & 0 & 0 & 0 & 0 & 1 & 0 \\
		 	0 & 0 & 1 & 0 & 0 & 0 & 0 & 0 \\
		 	0 & 0 & 0 & 0 & 0 & 0 & 0 & 1 \\
		 	0 & 0 & 0 & 1 & 0 & 0 & 0 & 0 \\
		 	0 & 0 & 0 & 0 & 0 & 1 & 0 & 0 \\
		 \end{array}
		 \right) \, .
		 \end{equation}
		 The first two lines and columns refer to the Cartan directions.
		 The inverse Killing metric is:
		 \begin{equation}
		 \kappa^{ab} = \left(
		 \begin{array}{cccccccc}
		 \frac{2}{3} & \frac{1}{3} & 0 & 0 & 0 & 0 & 0 & 0 \\
		 \frac{1}{3} & \frac{2}{3} & 0 & 0 & 0 & 0 & 0 & 0 \\
		 0 & 0 & 0 & 0 & 1 & 0 & 0 & 0 \\
		 0 & 0 & 0 & 0 & 0 & 0 & 1 & 0 \\
		 0 & 0 & 1 & 0 & 0 & 0 & 0 & 0 \\
		 0 & 0 & 0 & 0 & 0 & 0 & 0 & 1 \\
		 0 & 0 & 0 & 1 & 0 & 0 & 0 & 0 \\
		 0 & 0 & 0 & 0 & 0 & 1 & 0 & 0 \\
		 \end{array}
		 \right) \, .
		 \end{equation}
		 \subsubsection*{The Ten}
		 \label{su3canonicalrepresentatives}
		 An intermediate result in the calculation of the invariant $I_{10}$ in subsection \ref{SYMExample} is the determination of the $10$ representation at the second level of the exterior algebra $\Lambda \mathfrak{su}(3)$. The ten-dimensional representation is spanned by the following states:
		 \begin{eqnarray}
		 \psi_\theta \psi_{\alpha_1} \, , \quad \psi_{-\alpha_2} \psi_{-\theta} \, , \quad  \psi_{-\alpha_1} \psi_{\alpha_2} \, , \quad \frac{1}{\sqrt{3}} (\psi_{\alpha_2} \psi_{\alpha_1}+\psi_1 \psi_\theta) & & \nonumber \\
		   \frac{1}{\sqrt{3}} (\psi_{-\alpha_1} \psi_\theta+\psi_1 \psi_{\alpha_2}) 
		   \, , \quad  \frac{1}{\sqrt{3}} (\psi_{-\theta} \psi_{\alpha_2}+\psi_2 \psi_{-\alpha_1})  \, , \quad  \frac{1}{\sqrt{3}} (\psi_{-\alpha_2} \psi_{-\alpha_1}+\psi_2 \psi_{-\theta}) & & \nonumber \\
		  & &  \nonumber \\
		    \frac{1}{\sqrt{3}} (\psi_{\alpha_1} \psi_1 + \psi_{\alpha_1} \psi_2 + \psi_\theta \psi_{-\alpha_2} )
		    \, , \quad 
		      \frac{1}{\sqrt{3}} (\psi_{\alpha_1} \psi_{-\theta} + \psi_{-\alpha_2} \psi_1 + \psi_{-\alpha_2} \psi_{2} )
		    & & \nonumber \\
		      \frac{1}{\sqrt{6}} (\psi_{\alpha_1} \psi_{-\alpha_1} + \psi_{\alpha_2} \psi_{-\alpha_2} + \psi_{-\theta} \psi_\theta+ \psi_2 \psi_1  ) \, . & & 
		 \end{eqnarray}
		 
		\section{The  Coset Model and the Morse Potential}	
		\label{Coset}
		In this appendix, we define and briefly analyze the physical model whose ground states correspond to the $G/H$ coset de Rham cohomology, where $G$ is a connected compact group with simple Lie algebra $\mathfrak{t}$ (namely the compact form of $\mathfrak{g}$) and $H$ is a maximal torus with Lie algebra $\mathfrak{h}$. The analysis is a a straightforward  combination of the link between supersymmetric quantum mechanics and Morse theory \cite{Witten:1982im},
		and the identification of the relevant Morse function in the mathematics literature \cite{Bott,BottSamelson,Reeder}.
		
		We consider a supersymmetric quantum mechanics model with target space equal to the coset $G/H$. The variables of the model are the coordinates $\phi^a$ on the target as well as the tangent space fermions $\psi^a$. A given configuration has action \cite{Witten:1982im}: 
		\begin{equation}
		S = \frac{1}{2} \int d \tau \gamma_{ab}  (\partial_\tau \phi^a \partial_\tau \phi^b + i \bar{\psi}^a  D_\tau \psi^b) + \frac{1}{4} R_{abcd} \bar{\psi}^a \psi^c \bar{\psi}^b \psi^d
		- t^2 \gamma^{ab} \partial_a h \partial_b h - t D_a D_b h \bar{\psi}^a \psi^b \, ,
		\end{equation}
		where $\gamma_{ab}$ is the bi-invariant metric on the group, $R_{abcd}$ is the associated Riemann curvature tensor and we have introduce covariant derivatives.
		The action $S$ contains a superpotential function $h$ which will be the Morse function  \cite{Witten:1982im}, and a coupling constant $t$.
		After canonical quantization of the action the fermions acting on the vacuum generate the space of differential forms  \cite{Witten:1981nf}. 
		The number of zero eigenvalues of the Hamiltonian correspond one-to-one to the Betti numbers, at each degree of the space of differential forms, and the number of zero eigenvalues is independent of the coupling constant $t$  \cite{Witten:1982im}. For large coupling constant $t$, all ground state wave functions are concentrated at the critical points of the Morse function $h$ and with a proper choice of Morse function, it is possible to count them.

				 To define the Morse function $h$, we take a regular element $h_0$ of the Lie algebra  $\mathfrak{h}$. The Ad(G) centralizer of the regular element $h_0$ is the maximal torus $H$. We have that the coset $G/H$ is the $Ad(G)$ orbit of the element $h_0$. Choose now the Morse function $h$ on the coset to evaluate to \cite{Bott,BottSamelson,Reeder}:
			\begin{equation}
			h(gH) =  \langle \text{Ad}(g) h_0, h_0 \rangle \, .
			\end{equation}
			The derivative of the Morse function in the Lie algebra direction $X$ equals:
			\begin{equation}
			{X} h(gH) = \langle \text{Ad}(g) h_0, [h_0,X] \rangle \, .
			\end{equation}
				Recall the orthogonal decomposition $\mathfrak{t}=\mathfrak{m} \oplus \mathfrak{h}$ of the (real) simple Lie algebra in terms of the Cartan algebra 
			$\mathfrak{h}$ of the torus $H$ and its complement $\mathfrak{m}$. 		
   	We have that $\mathfrak{m}=\text{Im} \,  \text{ad}(h_0)$. Therefore the coset $gT$ is a critical point of the Morse function $h$ if and only if $\langle \text{Ad}(g) h_0, \mathfrak{m} \rangle $ =0 and therefore $\text{Ad}(g) h_0 \in \mathfrak{h}$.
   	We conclude that  we must have $
			\text{Ad}(g) h_0 = \text{Ad}(w) h_0$ for an element $w$ of the Weyl group $W$. 
			
			Moreover, the Hessian at each such coset point $wH$ can be computed in an orthonormal basis $X_a$ as follows \cite{Bott,BottSamelson,Reeder}:
			\begin{equation}
			h_{ab} (w) = X_a X_b h(wT) = \langle [ X_a,\text{Ad}(w)	 h_0], [ h_0,X_b ] \rangle  \, .
			\end{equation}
			This equals zero for $a \neq b$ and on the diagonal we have
			\begin{equation}
			h_{aa} = - \alpha_a(\text{Ad}(w) h_0) \alpha_a(h_0) 
			\, ,
			\end{equation}
			where we pick the generators $X_a$ to have weights equal to the roots $\alpha_a$.
			The Hessian is non-singular by the regularity of the Lie algebra element $h_0$, which implies also the regularity of the element $\text{Ad}(w) h_0$. The index $m$, or number of negative eigenvalues of the Hessian, is equal to twice the number of positive roots such that $w^{-1} \alpha$ is also a positive root. It is crucial that this number is always even. Morse theory than implies that  the Poincar\'e polynomial of the cohomology of the coset space $G/H$ is \cite{Bott,Witten:1981nf}:
			\begin{equation}
			P_{G/H}(u) = \sum_{w \in W} u^{2 m(w)} \, . \label{PoincarePolynomialCoset}
			\end{equation}
			The odd cohomology is zero. The even cohomology therefore has total dimension equal to the Euler number, which  equals  the number of elements in the Weyl group $W$. Therefore the relative cohomology has the same dimension as the space of harmonic polynomials. We note that it is possible in this context as well to determine the structure of the coset cohomology 
			 $H(G/H)$ as a Weyl group module. Since there is no odd cohomology, the Lefschetz number of a group element $w$ of the Weyl group equals its trace. If $w \neq 1$ then there are no fixed points and the trace is  zero. When $w=1$, it is the cardinal number $|W|$ of the Weyl group. This is precisely as in the regular representation of the Weyl group, in agreement with the description in the bulk of the paper in terms of harmonic polynomials. The coincidence between the Poincar\'e polynomial (\ref{PoincarePolynomialHarmonicPolynomials}) for the harmonic polynomials (with doubled degree) and the coset Poincar\'e polynomial (\ref{PoincarePolynomialCoset}) is yet another  non-trivial fact. See e.g.  \cite{Bott,SolomonChevalley} for proofs.

	\section{The  Free Fermion Hamiltonian}
	\label{PhysicsKumarProof}
	In this appendix, we provide a version of Kumar's proof \cite{KumarBook} of Garland's theorem \cite{Garland} that is readily accessible to physicists. The proof only  uses Wick's theorem for free fermions and elementary  algebra. The demonstration is rather long, but no longer than the known proofs in mathematics \cite{KumarBook}. Although this may not be manifest, we did use abstract mathematics \cite{KumarBook} to organize the calculation. We insert footnotes that compare our calculation to the mathematics proof in order to clarify this point -- the actual calculation however does not depend on the comparison.
	
	{From} the bulk of the paper, we recall the definition of the supercharges $\partial$ and $\partial^\dagger$  and the Hamiltonian $\Delta$:\footnote{The expression for the supercharges incorporates lemma 3.4.3 of \cite{KumarBook}. Lemma 3.4.4 is trivially satisfied for the  vacuum $|0\rangle $ that we have chosen.}
	\begin{eqnarray}
	\partial &=& \frac{1}{2} {f_a}^{bc} \psi^a_{-m-n} \psi^\dagger_{b,m} \psi^\dagger_{c,n}
	\nonumber \\
	\partial^\dagger &=&  \frac{1}{2} { f_{ab}}^c \psi^a_{-n} \psi^b_{-m} \psi^\dagger_{c,n+m}
	\nonumber \\
	\Delta  &=&  \partial \partial^\dagger +  \partial^\dagger \partial  \, .
	\end{eqnarray}
	We split (four times) the Hamiltonian $4 \Delta=T_1+T_2$ into two terms:
	\begin{eqnarray}
	T_1 &=& 4 \partial \partial^\dagger =  {f_d}^{ef} \psi^d_{-k-l} \psi^\dagger_{e,k} \psi^\dagger_{f,l}  { f_{ab}}^c \psi^a_{-n} \psi^b_{-m} \psi^\dagger_{c,n+m} 
	\nonumber \\
	T_2 &=& 4 \partial^\dagger \partial = { f_{ab}}^c \psi^a_{-n} \psi^b_{-m} \psi^\dagger_{c,n+m}  {f_d}^{ef} \psi^d_{-k-l} \psi^\dagger_{e,k} \psi^\dagger_{f,l} \, .
	\end{eqnarray}
	To compute the anti-commutator of the supercharges $\partial$ and $\partial^\dagger$, it is useful to know how to pull the $\psi^a_{-n}$ 
	factor through the operator $\partial$ in the term $T_1$.	
	We study the miniature version:
	\begin{eqnarray}
	{f_d}^{ef} \psi^d_{-k-l} \psi^\dagger_{e,k} \psi^\dagger_{f,l} \psi^a_{-n} &=& -\psi^a_{-n}  {f_d}^{ef} \psi^d_{-k-l} \psi^\dagger_{e,k} \psi^\dagger_{f,l}
	+ 2 {f_d}^{ea} \psi^d_{-k-n} \psi^\dagger_{e,k} \, .
	\end{eqnarray}
	We took into account two single contractions. This is the calculation of  the anti-commutator of the operator $\partial $ and exterior multiplication. We apply this anti-commutator  to the term $T_1$ and  find:\footnote{This is the content of equation (3.4.5.3) in \cite{KumarBook}.}
	\begin{eqnarray}
	T_1 &=& T_{11}+T_{12}
	\nonumber \\
	T_{11} &=& -\psi^a_{-n}  {f_d}^{ef} \psi^d_{-k-l} \psi^\dagger_{e,k} \psi^\dagger_{f,l} { f_{ab}}^c  \psi^b_{-m} \psi^\dagger_{c,n+m} 
	\nonumber \\
	T_{12} &=& 2 {f_d}^{ea} \psi^d_{-k-n} \psi^\dagger_{e,k}  { f_{ab}}^c \psi^b_{-m} \psi^\dagger_{c,n+m} \, .
	\end{eqnarray}	 
	We concentrate on the term $T_{12}$ first. 
	One manner to study the term $T_{12}$ further is to  act on an explicit decomposable state $\Psi=\psi_1 \psi_2 \dots \psi_k \in \Lambda^k(\mathfrak{u}^-)$. The operator $T_{12}$ can either act on two different factors, or on twice the same factor. We thus have:
	\begin{eqnarray}
	T_{12} \Psi &=& \sum_{s=1}^k (-1)^{s-1} (T_{12} \psi_s) \Psi^s
	+ 2 \sum_{s=1}^p (-1)^{s-1} ({ f_{ab}}^c \psi^b_{-m} \psi^\dagger_{c,n+m} \psi_s)  {f_d}^{ea} \psi^d_{-k-n} \psi^\dagger_{e,k} \Psi^s \, .
	\nonumber \\
	&=& T_{121} ^\Psi+ T_{122} ^\Psi \, . \label{T12split}
	\end{eqnarray}
	We have defined the state $\Psi^s = \psi_1  \dots  \hat{\psi_s}  \dots  \psi_p$ which is missing the factor  $\psi_s$, and we have given names to the first and second terms in equation (\ref{T12split}).
	The first $T_{121}^\Psi$ piece will be part of the end result.
	The second  $T_{122}^\Psi$ term
	is to be canceled at a later stage of the calculation. 
	
	Thus, we leave aside the $T_{12}$ term in the term $T_1$ in the Hamiltonian $4 \Delta$, and come back to the piece $T_{11}$. For simplicity, we temporarily strip off   the factor of $-\psi^a_{-n}$ on the left, and analyze the  operator $\partial$ that makes up the left factor.
	We trivially rename and lower and raise indices to define the resulting operator $S^a$:
	\begin{eqnarray}
	S^a &=& {f^d}_{bc} \psi_{d,-k-m} \psi_k^{\dagger b} \psi_m^{\dagger c} {f^a}_{ef} \psi^e_{-l} \psi^{\dagger f}_{l+n} \, .
	\end{eqnarray}
	For notational simplicity, we  drop the upper index $a$ on the operator $S$ in the following. Our calculation proceeds by
	firstly pulling the $\psi^\dagger$ annihilation operators to the right:
	\begin{eqnarray}
	S &=& S_1 + S_2
	\nonumber \\
	S_1 &=& {f^d}_{bc}     \psi_{d,-k-m}  {f^a}_{ef} \psi^e_{-l} \psi^{\dagger f}_{l+n} \psi_k^{\dagger b} \psi_m^{\dagger c} 
	\nonumber \\
	S_2 &=&  2 {f^d}_{bc} \psi_{d,-k-m} \psi_k^{\dagger b}  {f^a}_{cf}  \psi^{\dagger f}_{m+n} \, .
	\end{eqnarray}
	The term $S_2$ results from two single contractions.
	In the term
	$S_1$ only, we pull  the $\psi_{d}$ factor through the $\psi^{\dagger f}$ operator to obtain the split:
	\begin{eqnarray}
	S_1 &=& S_{11}+ S_{12}
	\nonumber \\
	S_{11}&=& {f^d}_{bc}      {f^a}_{ef} \psi^e_{-l} \psi^{\dagger f}_{l+n}  \psi_{d,-k-m}  \psi_k^{\dagger b} \psi_m^{\dagger c} 
	\nonumber \\
	S_{12}&=& -{f^d}_{bc}      {f^a}_{ed} \delta_{l+n-k-m} \psi^e_{-l}  \psi_k^{\dagger b} \psi_m^{\dagger c} \, .
	\end{eqnarray}
	The term $S_{11}$ where no contractions were performed  cancels the term $T_2$ (after restoring the $\psi^a_{-n}$ factor).  We analyze the sum of the terms $S_{12}$ and $S_2$, namely the three single contraction terms, by  evaluating them explicitly on a decomposable vector $\Psi$ of degree $k$ which we moreover pick to be:
	\begin{eqnarray}
	\Psi &=& \psi_1   \dots   \psi_k
	= \psi^{a_1}_{-n_1} \dots  \psi^{a_k}_{-n_k} |0 \rangle \, . \label{ParticularPsi}
	\end{eqnarray}
	We evaluate:
	\begin{eqnarray}
	(S_{12}+S_2) \Psi &=&  - ( \sum_{s>s'} - \sum_{s<s'}) (-1)^{s+s'} {f^d}_{a_{s'} a_s}      {f^a}_{ed} \delta_{l+n-n_{s'}-n_s} \psi^e_{-l} 
	\Psi^{(s,s')}  \nonumber \\
	&&
	+ 2  ( \sum_{s>s'} - \sum_{s<s'}) (-1)^{s+s'}  {f^d}_{a_{s'} c} \psi_{d,-n_{s'}-m}   {f^a}_{c a_s} \delta_{m+n-n_s} \Psi^{(s,s')}  \, ,
	\end{eqnarray}
	where we used the notation $\Psi^{(s,s')}$ to denote the state vector $\Psi$ with $\psi_s$ and $\psi_{s'}$ factors omitted.
	We concentrate on rewriting the first term, using the Jacobi identity and relabelling:
	\begin{eqnarray}
	S_{12} \Psi &=&  ( \sum_{s>s'} - \sum_{s<s'}) (-1)^{s+s'}  ( {f^d}_{a_s e} {f^a}_{a_{s'} d} +  {f^d}_{e a_{s'}} {f^a}_{a_{s} d}        )   \delta_{l+n-n_{s'}-n_s} \psi^e_{-l} 
	\Psi^{(s,s')} 
	\nonumber \\
	&=& 2  ( \sum_{s>s'} - \sum_{s<s'}) (-1)^{s+s'}    {f^d}_{e a_{s'}} {f^a}_{a_{s} d}          \delta_{l+n-n_{s'}-n_s} \psi^e_{-l}  \Psi^{(s,s')} \, .
	\end{eqnarray}
	Thus, for the sum, we obtain:
	\begin{eqnarray}
	(S_{12}+S_2) \Psi &=&  2  ( \sum_{s>s'} - \sum_{s<s'}) (-1)^{s+s'}    {f^d}_{e a_{s'}} {f^a}_{a_{s} d}          \delta_{l+n-n_{s'}-n_s} \psi^e_{-l}   \Psi^{(s,s')} \nonumber \\
	&&
	+ 2  ( \sum_{s>s'} - \sum_{s<s'}) (-1)^{s+s'}  {f}_{e a_{s'} d} \psi^e_{-n_{s'}-m}   {f^a}_{d a_s} \delta_{m+n-n_s} \Psi^{(s,s')}
	\nonumber \\
	&=& 2  ( \sum_{s>s'} - \sum_{s<s'}) (-1)^{s+s'}    {f^d}_{e a_{s'}} {f^a}_{a_{s} d}  
	(  \delta_{l+n-n_{s'}-n_s} \psi^e_{-l}  - \delta_{m+n-n_s} \psi^e_{-n_{s'}-m})  \Psi^{(s,s')}
	\, . \nonumber 
	\end{eqnarray}
	We need to carefully analyze the sum ranges in order to simplify the expression. We have that the first term is summed over the level $l$ and that the level equals
	$l=n_s+n_{s'}-n$. Since we have the inequality $l \ge 1$ we must have that $n_s+n_{s'}-n \ge 1$. This is the only constraint on the first term. In the second term, we have the equality
	$m = n_s-n$ between levels and since the inequality $m \ge 1$ holds, we have that $n_s-n \ge 1$. Since $n_{s'} \ge 1$, this constraint is strictly stronger. Thus, after the subtraction, we are left with the terms:
	\begin{eqnarray}
	(S_{12}+S_2) \Psi &=&  2  ( \sum_{s>s'} - \sum_{s<s'}) (-1)^{s+s'}    {f^d}_{e a_{s'}} {f^a}_{a_{s} d}  
	\psi^e_{-n_s-n_{s'}+n}     \Psi^{(s,s')} \theta(n-n_s) 
	\, . \label{intermediate}
	\end{eqnarray}	
	The $\theta$-function equals $1$ for zero argument.  The further constraint $n_s+n_{s'}-n \ge 1$ is still implied by the index notation in equation (\ref{intermediate}).\footnote{The final expression is the equivalent of 
		equation (3.4.5.5) in \cite{KumarBook}.}  We now split the term (\ref{intermediate}) into two pieces. One piece corresponds to the value $n=n_s$ and the other has $n-n_s \ge 1$. For the latter piece, we
	set $n-n_s = m \ge 1$.  
	We write:
	\begin{eqnarray}
	(S_{12}+S_2) \Psi &=&  2  ( \sum_{s>s'} - \sum_{s<s'}) (-1)^{s+s'}    {f^d}_{e a_{s'}} {f^a}_{a_{s} d}  
	\psi^e_{-n_{s'}}     \Psi^{(s,s')} \nonumber \\
	& & 	+  2  ( \sum_{s>s'} - \sum_{s<s'}) (-1)^{s+s'}    {f^d}_{e a_{s'}} {f^a}_{a_{s} d}  
	\psi^e_{m-n_{s'}}   \delta_{m-n+n_s}  \Psi^{(s,s')} 
	\, . \label{intermediateTwo}
	\end{eqnarray}	
	{From} the second term in equation (\ref{intermediateTwo}) we obtain (after restoring the factor $-\psi^a_{-n}$)
	\begin{eqnarray}
	-\psi^{a}_{-n}	(S_{12}+S_2)_2 \Psi &=&  - 2  ( \sum_{s>s'} - \sum_{s<s'}) (-1)^{s+s'}    {f^d}_{e a_{s'}} {f}_{a a_{s} d}  \psi^{a}_{-n}
	\psi^e_{m-n_{s'}}    \delta_{m-n+n_s}  \Psi^{(s,s')} \nonumber \\
	&=&   -2  ( \sum_{s>s'} - \sum_{s<s'}) (-1)^{s+s'}    {f^a}_{d a_{s'}} {f}_{b a_{s} a}  \psi^{b}_{-m}
	\psi^d_{n-n_{s'}}    \delta_{n-m+n_s}  \Psi^{(s,s')} 
	\,  \label{S12S2secondterm}
	\end{eqnarray}
	Recall now that we defined:
	\begin{eqnarray}
	T_{122} \Psi &=& 2 \sum_{s=1}^k (-1)^{s-1} ({ f_{ab}}^c \psi^b_{-m} \psi^\dagger_{c,n+m} \psi_{-n_s}^{a_s})  {f_d}^{ea} \psi^d_{-k-n} \psi^\dagger_{e,k} \Psi^s 
	\nonumber \\
	&=& 2 \sum_{s=1}^k (-1)^{s-1} { f_{ab}}^{a_s} \psi^b_{n-n_s} {f_d}^{ea} \psi^d_{-k-n} \psi^\dagger_{e,k} \Psi^s 
	\nonumber \\
	&=&2 (\sum_{s>s'}-\sum_{s<s'}) (-1)^{s+s'} { f_{ab}}^{a_s} \psi^b_{n-n_s} {f_d}^{a_{s'}a} \psi^d_{-n_{s'}-n}  \Psi^{(s,s')} 
	\nonumber \\
	&=&2 (\sum_{s>s'}-\sum_{s<s'}) (-1)^{s+s'} { f_{ad}}^{a_{s'}} {f_b}^{a_{s}a}  \psi^b_{-n_{s}-n} \psi^d_{n-n_{s'}}  \Psi^{(s,s')}  \, . \label{T122term}
	\end{eqnarray}
	where we plugged in $\psi_s=\psi_{-n_s}^{a_s}$.
	%
	We note that the term $T_{122} \Psi$ in equation (\ref{T122term}) cancels the term recorded in equation (\ref{S12S2secondterm}). We are still left with the first term in (\ref{intermediateTwo}) as well as the term
	$T_{121}$ in equation (\ref{T12split}).\footnote{By now, we have proven Lemma 3.4.5 in \cite{KumarBook}.}
	Thus, we summarize:
	\begin{eqnarray}
	4 \Delta (\Psi) &=& \sum_{s=1}^p (-1)^{s-1} (T_{12} \psi_{-n_s}^{a_s}) \Psi^s - \psi_{-n_s}^a (S_{12}+S_1)_1 \, .
	\end{eqnarray}
	We compute:
	\begin{eqnarray}
	T_{12} \psi_{-n_s}^{a_s} &=& 2 {f_d}^{ea} \psi^d_{-k-n} \psi^\dagger_{e,k}  { f_{ab}}^c \psi^b_{-m} \psi^\dagger_{c,n+m}  \psi_{-n_s}^{a_s}
	\nonumber \\
	&=&  2  \psi^{a_s}_{-n_s}   \delta_{n_s,n+k} = 2 (n_s-1) \psi^{a_s}_{-n_s} \, ,
	\end{eqnarray}
	where we used the inequalities $n,k \ge 1$.
	When we put everything together, we finally find:
	\begin{eqnarray}
	4 \Delta (\Psi) 	&=& 2 \sum_s (n_s-1) \Psi  -2  ( \sum_{s>s'} - \sum_{s<s'}) (-1)^{s+s'}    {f^d}_{e a_{s'}} {f}_{a a_{s} d}  
	\psi_{-n_s}^a \psi^e_{-n_{s'}}     \Psi^{(s,s')}  \nonumber \\
	&=& 2 \sum_s (n_s-1) \Psi +   {f^d}_{e g} {f^b}_{c d}  \psi^e_0 \psi^{\dagger, g}_0 \psi_{0,b} \psi_0^{\dagger c} \Psi
	- \sum_s (-1)^{s-1} {f^d}_{e g} {f^b}_{c d}  \psi^e_0 \psi^{\dagger, g}_0 \psi_{0,b} \psi_0^{\dagger c} \psi_s \wedge \Psi^{(s)}
	\nonumber \\
	&=&2 \sum_s n_s \Psi +   ({f^d}_{e g} {f^b}_{c d}  \psi^e_0 \psi^{\dagger, g}_0 \psi_{0,b} \psi_0^{\dagger c}) \Psi
	\nonumber \\
	&=& 2 \sum_s (n_s-1) \Psi -2 f_{da b} \psi^a_{-n} \psi^{\dagger b}_n {f^d}_{e c} \psi^e_{-m} \psi^{\dagger c}_m ( \Psi) +2 s \Psi
	\nonumber \\
	&=& 2 (\sum_s n_s  - f_{da b} \psi^a_{-n} \psi^{\dagger b}_n {f^d}_{e c} \psi^e_{-m} \psi^{\dagger c}_m) ( \Psi) 
	\nonumber \\
	&=& 2 (L-C_2) \Psi \, .
	\end{eqnarray}
	In the last line, we defined the operator $L$ that measures the level of a state $\Psi$:
	\begin{equation}
	L \psi^{a_1}_{-n_1} \dots \psi^{a_k}_{-n_k} | 0 \rangle = (\sum_{s=1}^k {n_s}) \psi^{a_1}_{-n_1} \dots \psi^{a_k}_{-n_k} | 0 \rangle \, ,
	\end{equation}
	as well as the quadratic Casimir operator acting on the Hilbert space $\Lambda \mathfrak{u}^-$:
	\begin{equation}
	C_2 = f_{da b} \psi^a_{-n} \psi^{\dagger b}_n {f^d}_{e c} \psi^e_{-m} \psi^{\dagger c}_m \, .
	\end{equation}
	This is the quadratic Casimir for the adjoint action of the algebra $\mathfrak{g}$ on the Hilbert space. Thus, if we have an irreducible $\mathfrak{g}$ representation component of $\Lambda \mathfrak{u}^-$ with highest weight $\mu$ and level $N$, then the Laplacian equals:
	\begin{equation}
	\Delta = \frac{1}{2} ( N - (\rho+\mu,\rho+\mu)+(\rho,\rho)) \, ,
	\end{equation}
	on that irreducible component.\footnote{This corresponds to the affine case of theorem 3.4.2 in \cite{KumarBook}, for the trivial representation for the fermion zero modes.}

	\bibliographystyle{JHEP}

\end{document}